\documentclass[aps,10pt,pra,twocolumn,superscriptaddress,floatfix]{revtex4-2}

\usepackage{graphicx}
\usepackage{amssymb}
\usepackage{makecell}
\usepackage{float}
\usepackage{amsmath}
\usepackage{amsthm}
\usepackage{threeparttable}
\usepackage{bm}
\usepackage[colorlinks,
linkcolor=blue,      
anchorcolor=pink, 
citecolor=brown]{hyperref}
\usepackage{color,xcolor}
\usepackage{subfigure}
 \usepackage{algpseudocode}
\usepackage{soul}
\usepackage{algorithm}
\usepackage{algorithmicx}
 \usepackage{lineno}
\usepackage{soul,xcolor}
\usepackage{color}
\usepackage{multirow}
\usepackage{hhline}
\usepackage{booktabs}

\usepackage{rotating}
\usepackage{tabularray}
\usepackage{listings}
\usepackage{xcolor}
\definecolor{codegreen}{rgb}{0,0.6,0}
\definecolor{codegray}{rgb}{0.5,0.5,0.5}
\definecolor{codepurple}{rgb}{0.58,0,0.82}
\definecolor{backcolour}{rgb}{0.95,0.95,0.92}

\lstdefinestyle{mystyle}{
    backgroundcolor=\color{backcolour},   
    commentstyle=\color{codegreen},
    keywordstyle=\color{magenta},
    numberstyle=\tiny\color{codegray},
    stringstyle=\color{codepurple},
    basicstyle=\ttfamily\footnotesize,
    breakatwhitespace=false,         
    breaklines=true,                 
    captionpos=b,                    
    keepspaces=true,                 
    numbers=left,                    
    numbersep=5pt,                  
    showspaces=false,                
    showstringspaces=false,
    showtabs=false,                  
    tabsize=2
}

\begin{document}
\title{Quantum neural compressive sensing for ghost imaging}
	
\author{Xinliang Zhai}	\affiliation
{State Key Laboratory of Advanced Optical Communication Systems and Networks, and Institute for Quantum Sensing and Information Processing, Shanghai Jiao Tong University, Shanghai 200240, China
}

\author{Tailong Xiao}
\thanks{Xinliang Zhai and Tailong Xiao contributed equally}
\email{tailong\_shaw@sjtu.edu.cn}
\affiliation
{State Key Laboratory of Advanced Optical Communication Systems and Networks, and Institute for Quantum Sensing and Information Processing, Shanghai Jiao Tong University, Shanghai 200240, China
}
\affiliation{Hefei National Laboratory, Hefei 230088, China}
\affiliation{Shanghai Research Center for Quantum Sciences, Shanghai 201315, China}

\author{Jingzheng Huang}\affiliation
{State Key Laboratory of Advanced Optical Communication Systems and Networks, and Institute for Quantum Sensing and Information Processing, Shanghai Jiao Tong University, Shanghai 200240, China
}\affiliation{Hefei National Laboratory, Hefei 230088, China}\affiliation{Shanghai Research Center for Quantum Sciences, Shanghai 201315, China}

\author{Jianping Fan}\affiliation
{AI Lab, Lenovo Research, Beijing 100094, China}

\author{Guihua Zeng}\email{ghzeng@sjtu.edu.cn}\affiliation
{State Key Laboratory of Advanced Optical Communication Systems and Networks, and Institute for Quantum Sensing and Information Processing, Shanghai Jiao Tong University, Shanghai 200240, China
}\affiliation{Hefei National Laboratory, Hefei 230088, China}\affiliation{Shanghai Research Center for Quantum Sciences, Shanghai 201315, China}
\date{\today}

\begin{abstract}
Demonstrating the utility of quantum algorithms is a long-standing challenge, where quantum machine learning becomes one of the most promising candidate that can be resorted to. In this study, we investigate a quantum neural compressive sensing algorithm for ghost imaging to showcase its utility. The algorithm utilizes the variational quantum circuits to reparameterize the inverse problem of ghost imaging and uses the inductive bias of the physical forward model to perform optimization. To validate the algorithm's effectiveness, we conduct optical ghost imaging experiments, capturing signals from objects at different physical sampling rates and detection signal-to-noise ratios. The experimental results show that our proposed algorithm surpasses conventional methods in both visual appearance and quantitative metrics, achieving state-of-the-art performance. Importantly, we observe that the quantum neural network, guided by prior knowledge of physics, effectively overcomes the challenge of barren plateau in the optimization process. The proposed algorithm demonstrates robustness against various quantum noise levels, making it suitable for near-term quantum devices. Our study leverages physical inductive bias guided variational quantum algorithm, underscoring the potential of quantum computation in tackling a broad range of optimization and inverse problems.
\end{abstract}
\maketitle


\section{Introduction}
Quantum hardware has entered into 1,000 qubits era, and the fidelity of the quantum operations has also been improved rapidly. The quantum utility of using quantum computers has also been demonstrated in various physical platforms. However, these quantum algorithms such as random circuit sampling has limited utility in practical applications. Exploiting the practical utility of quantum algorithms becomes the goal in noisy intermediate-scale quantum (NISQ) devices.

Concurrently, quantum machine learning (QML) has garnered considerable attention for its efficiency and potential for stronger expressive power \cite{biamonte2017quantum, schuld2015introduction, caro2022generalization,xiao2023quantum,xiao2022parameter,xiao2022intelligent}. Even in the NISQ era \cite{preskill2018quantum}, QML continues to hold advantages over classical machine learning in terms of generalization \cite{cerezo2022challenges, banchi2021generalization, peters2023generalization, gibbs2024dynamical, caro2023out} and learning ability \cite{schuld2021effect,du2020expressive,abbas2021power,du2020expressive,jerbi2023quantum}. These include applications like the quantum generative adversarial network \cite{lloyd2018quantum, niu2022entangling, dallaire2018quantum, huang2021experimental} and quantum circuit Born machine \cite{gili2023quantum, liu2018differentiable, benedetti2019parameterized}. The potential advantage is that quantum machine learning maps the data into the quantum Hilbert feature space thus leading to a better pattern recognition ability \cite{schuld2019quantum}. Ref. \cite{xiao2023practical} found that quantum machine learning has practical advantage over classical machine learning in reduced ghost imaging (GI) scenarios in terms of sample complexity. 

The utility of ghost imaging (GI) is constrained by both the number of samplings and the quality of reconstruction, posing a pressing challenge that demands efficient imaging algorithms. \textcolor{black}{ As a typical computational imaging scheme, GI is based on a sequence of one-dimensional bucket signals acquired under modulated illumination for image reconstruction \cite{gatti2004ghost, ryczkowski2016ghost, wang2023photon, liu2017single, li2022ghost}.}
Existing GI reconstruction approaches can be categorized into three main groups: iteration-based methods, model-based methods, and learning-based methods. Iteration-based methods, such as differential ghost imaging (DGI) \cite{ferri2010differential}, rely on the correlation between bucket signals and corresponding illumination patterns. While simple, they struggle to produce high-quality images with a limited number of samplings. In practical applications, GI reconstruction often faces an underdetermined problem. Therefore, model-based methods utilize a priori information about the image, such as sparsity, in conjunction with the physical model of GI to optimize the reconstructed image. For instance, some researchers have applied compressive sensing algorithms to GI \cite{duarte2008single, zhao2012ghostlidar, katz2009compressive}. With the advent of deep learning, learning-based GI \cite{lyu2017deep, wang2019learning, rizvi2020deepghost} approaches have effectively addressed the trade-off between sampling numbers and imaging quality, although they require the collection of datasets to train deep learning models.
Recent developments propose using classical neural network to reparameterize the optimization process in GI, referred to as GI with untrained neural networks \cite{wang2022single, wang2022far}. However, the potential of untrained quantum neural network for GI has not been investigated.

In this study, we propose a quantum neural compressive sensing GI algorithm (QCSGI). This algorithm utilizes quantum neural networks (QNNs) as optimization constraints and integrates a physical forward model to compute the loss function, optimizing the hybrid QNN ansatz. Differing from methods like the variational quantum eigensolver (VQE) \cite{tilly2022variational} and other QNN-based machine learning algorithms \cite{martin2022quantum, xiao2023practical}, our proposed algorithm does not require datasets, ground-truth data, or reliance on the variational principle. Besides both combining GI and QNNs, the approach in this paper is quite different from our previous work \cite{xiao2023practical}. Ref. \cite{xiao2023practical} treats GI as a supervised learning task which needs pretraining, whereas the proposed QCSGI in this work treats the network as an iterative solver for compressive sensing based GI. It is clearly that these two methods are applicable to different application scenarios. To validate the reconstruction performance of QCSGI, we conducted physical experiments in a GI optical system to collect measurement data. The results demonstrate QCSGI's superiority in object recovery over classical neural networks and DGI methods. QNN-reconstructed images exhibit finer details compared to smoother convolutional neural network (CNN) recovered images. We also assess the algorithm's performance under noisy measurements, demonstrating its robustness in low signal-to-noise ratio scenarios.

Furthermore, we examine the impact of different quantum encoding-enabled circuits and quantum noise on the algorithm's reconstruction performance. Under varying quantum noise rates, QCSGI consistently achieves similar performance as in noise-free conditions. Additionally, we find that instant quantum polynomial (IQP) and Heisenberg-evolution encoding strategies yield competitive results compared to the angle encoding strategy in image recovery. Intriguingly, we calculate the gradient variance of randomly initialized quantum circuits under the loss function induced by the physical forward model, revealing the absence of the Barren Plateaus problem. This result is likely due to two reasons: 1) classical neural networks can reshape the loss landscape to mitigate the Barren Plateaus problem \cite{mcclean2018barren,friedrich2022avoiding, rivera2021avoiding}; 2) the loss landscape, induced by the physical forward model, incorporates inductive biases from the data and the experimental system, rendering it suitable for optimization using gradient descent methods. Our proposal, QCSGI, is the first algorithm, to our knowledge, that integrates the physical inductive bias into the optimization loop of QNNs, offering significant potential in compressed sensing, GI, and other physical applications.

The remainder of this work is structured as follows: Section \ref{Methods} introduces the methods, encompassing the imaging scheme, network design, and optimization techniques. In Section \ref{R-D}, we present the experimental results of the algorithm, including system configurations, outcomes under different sampling ratios, and varying signal-to-noise ratios (dSNRs). Finally, we offer concluding remarks in Section \ref{Summary} to summarize the key contributions of this study.

\section{Methods}\label{Methods}
\subsection{Quantum neural compressive sensing for GI}

\begin{figure*}[htbp]
	\centering
	\includegraphics[width=0.8\textwidth]{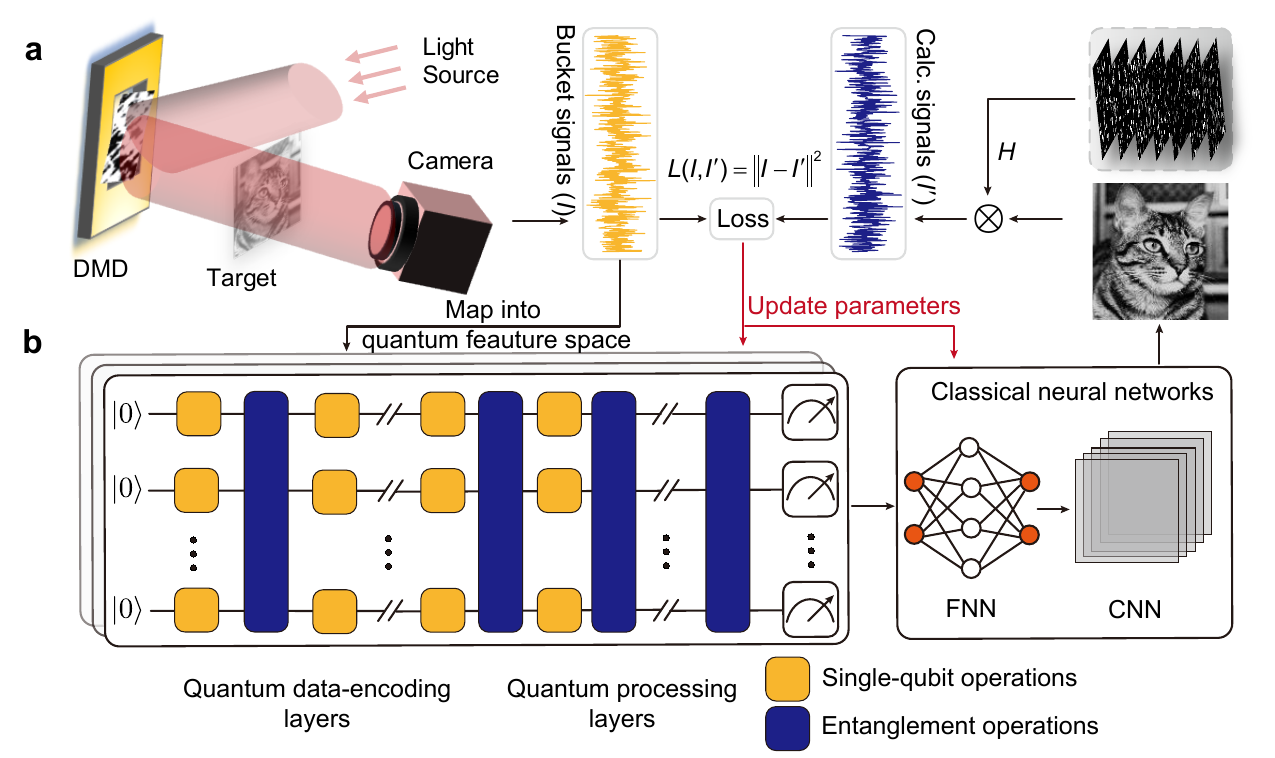}
	\caption{Compressive ghost imaging via a variational quantum circuit constraint. (a) The experimental schematic of ghost imaging where the light source is modulated by spatial light modulator with prescribed illumination patterns. The modulated light field is subsequently interacted with the target before being measured with a single-pixel detector. (b) The hybrid quantum neural network where the bucket signals firstly normalized into $[0,2\pi]$ are mapped into the feature space through a variational quantum circuit and then processed by fully-connected and convolution neural networks.  }
	\label{fig:1}
\end{figure*}

Compressive sensing (CS)\cite{candes2006compressive, baraniuk2007compressive, romberg2008imaging} is a signal processing technique that enables the recovery of sparse or compressible signals from a small number of non-adaptive linear measurements. It is particularly useful when the number of measurements is much smaller than what is typically required by the Nyquist-Shannon sampling theorem. Mathematically, we can express it as 
\begin{equation}\label{eq.0}
	\bm{S}^* = \arg\min\|\bm{S} \|\quad \text{s.t.} \quad \bm{I} = \bm{H}\bm{P}\bm{S},
\end{equation}
where $\bm{S}$ is  the sparse coefficients of original signal $O$ under the sparse basis $\bm{P}$, that is $\bm{O} = \bm{P}\bm{S}$. $\bm{I}$ is the measurements under measurement basis $\bm{H}$. The goal of compressive sensing is to reconstruct the original sparse signal $\bm{O}$ from the measurements $\bm{I}$. In our case of compressive ghost imaging, we require reconstructing the 2D image of the object from the detected 1D bucket signals under compressive sampling.

As shown in Fig. \ref{fig:1},  the proposed method is based on the typical computational GI system depicted in Fig. \ref{fig:1}a.  First, a reflective Digital Micromirror Device (DMD) is utilized as the spatial light field modulator. The output light field from the DMD is then interacting with the object. Finally, the object light is collected by a bucket detector. Thus, the forward model can be formulated as 
\begin{equation}\label{eq.1}
	I_j = \iint H_j(x, y)T(x, y)\text{d}x\text{d}y + n_j,
\end{equation}
where $T(x, y)$ is the intensity transmission function of object. $H_j(x, y)$ and $I_j$ denote the $j$-th DMD's modulation and the corresponding bucket detection, respectively. The noise term $n_j$ includes background noise and detection noise.  

According to GI theory \cite{pittman1995optical, bennink2002two, gatti2004ghost, shapiro2008computational}, the object's image $O_{\text{GI}}$ can be retrieved by correlating the measured intensity and the speckle field (measurement basis) $H_j$. 
\begin{equation}\label{eq.2}
	O_{\text{GI}}(x, y) = \frac{1}{M}\sum_{j=0}^{M-1} (I_j - \langle I \rangle)(H_j(x, y) -  \langle H(x, y) \rangle)
\end{equation}
where the number of samplings is $M$  and $\langle \cdot \rangle$ denotes the ensemble average for $M$ iterations. To reconstruct an $N$-pixel image, the sampling number $M$ needs to be at least equal to $N$ (in the orthogonal case). 
However, GI is always performed under $M \ll N$ for real-time imaging. Thus, the solution to the inverse process of Eq. \eqref{eq.1} is underdetermined, which is a classic optimization problem. However, the solution is highly sensitive to noise when the number of samplings is small.  To further improve the results, we can introduce the sparse prior information in retrieval algorithm according to Eq. \eqref{eq.0}. Therefore, the objective function can be expressed as 
\begin{equation}\label{eq.3}
	\bm{O}^*_{\text{CS}} = \arg\min\frac{1}{2}\|\bm{I} - \bm{H}{\bm{O}_{\text{CS}}}\|_2^2+\mu\|\bm{O}_{\text{CS}}\|_{\text{TV}},
\end{equation}
where the first term is a basic convex optimization problem and the second term is a sparsity constraint in the pixel domain weighted by the factor $\mu$. Specifically, a common choice for this constraint is total variation (TV) which is described as 
\begin{equation}\label{eq. 4}
	\|\bm{O}_{\text{CS}}\|_{\text{TV}} = \sum_{i=0}^{N-1} \left ( \frac{\partial  \bm{O}_{\text{CS}}}{\partial x} + \frac{\partial  \bm{O}_{\text{CS}}}{\partial y} \right ).
\end{equation}
Under the constraint of sparsity prior, CS can improve the imaging quality. However, when $M/N$ is less than the Cramer-Rao bound \cite{donoho2006compressed, eldar2012theory}, Eq. \eqref{eq.3} is still a challenging problem. 

A natural idea is to introduce more prior information into Eq. \eqref{eq.3}. According to deep image prior theory \cite{ulyanov2018deep}, a neural network itself
has an implicit bias towards natural images and performs better in solving ill-posed inverse problems of GI \cite{wang2022single, wang2022far}. Otherwise, compared with conventional neural networks, the quantum circuit shows more powerful learning capability and lighter network size. Thus, an untrained hybrid quantum-classical neural network can be used to reparameterize the optimization problem and we can rewrite the Eq. \eqref{eq.3} as 
\begin{equation}\label{eq.5}
	\bm{O}^* = \arg\min\frac{1}{2}\|{\bm{I}} - \bm{H}\Phi(\bm{z}, \bm{\theta})\|_2^2+\mu\|\Phi(\bm{z}, \bm{\theta})\|_{\text{TV}},
\end{equation}
where $\Phi$ is the proposed QNN with the input $\bm{z}$ and trainable parameters $\bm{\theta}$. The input of QNN is the bucket signal vector $\bm{I}$. \textcolor{black}{For high-dimensional bucket vectors with a large number of samples, a patch strategy can be employed to divide the high-dimensional vector into multiple low-dimensional bucket vectors. This allows for two approaches: either multiplexing a quantum neural network (QNN) $\Phi(\bm z, \bm \theta)$ to temporally learn the features, or using multiple parallel QNNs $\{\Phi_i(\bm z, \bm \theta)\}_i$ with a relatively smaller number of qubits to extract the features. A detailed explanation is provided in Appendix \ref{optimization}. The latter approach is more suitable for distributed quantum computing. Intuitively, the patch strategy preserves information since the high-dimensional bucket vector is generated under independent and identically distributed (i.i.d.) random patterns. Each bucket value is independent and can be separately mapped to a QNN for feature learning.}

As shown in Fig.~\ref{fig:1}, the proposed QCSGI consists of a QNN and a classical neural network. Firstly, QNN maps the normalized input $(\bm{z}\in [0,2\pi]^M)$ to a high-dimensional quantum feature space and learns a good feature representation \cite{xiao2023practical,schuld2019quantum}. Subsequently, a classical neural network is applied to the learned representation to reconstruct the final object image. The data pipeline is well suited for hybrid quantum-classical platforms \cite{ang2024arquin}. The detailed hybrid qunatum-classical neural network and its optimization algorithm can be found in Appendix \ref{optimization}. 

To solve Eq. (\ref{eq.5}), we optimize the QNN ansatz by the gradient descent method. Firstly, the loss function is calculated by the mean square error (MSE) of the experimental bucket signals and the estimated bucket signals given by
\begin{equation}
	\mathcal{L}(\bm{I},\bm{I}')
	= \sum_{i=1}^{M} \left(I_i - {H}_i\hat{\mathcal{Y}}\right)^2 + \mu \left\|\hat{\mathcal{Y}} \right\|_{\text{TV}},
\end{equation}
where $H_i$ denotes the $i$th random pattern used to modulate the light field, $\hat{\mathcal{Y}}$ denotes the estimated image by QCSGI, $\hat{\mathcal{Y}} = \Phi(\bm{z}, \bm{\theta})$, $\bm{I}'$ is the estimated bucket signal. The gradient of the loss function over the network can be divided two parts (see Appendix \ref{optimization}).
While the gradients of the loss function over the parameters are calculated, we can use the gradient descent method to update the parameters with learning rate $\alpha$,
\begin{equation}
	\bm\theta, \bm\nu \leftarrow \bm\theta,\bm\nu - \alpha \partial_{\bm\theta,\bm\nu} \mathcal{L}(\bm\theta,\bm\nu).
\end{equation}
Generally, we can use Adam \cite{kingma2014adam} optimizer to learn the quantum and classical parameters. Although QNN does not support AD, the forward speed is highly fast so that it can also calculate the gradients with the parameter-shift rule. Recently, Ref. \cite{bowles2023backpropagation} points out that QNN is also able to own a back-propagation scaling as the classical neural network does with a constrained quantum circuit structure, which may further be applied into our proposed QNN to enhance the learning efficiency. The procedure of the proposed QNN algorithm for compressive sensing GI is presented in the Algorithm \ref{algorithm}.

\begin{algorithm}[H]
	\caption{QCSGI}\label{algorithm}	
   \begin{algorithmic}[1]
   \State Input: total iteration time $T$, mean square error (MSE) threshold $\varepsilon$, gradient norm threshold $\epsilon$, initial MSE $\mathcal{L}$, experimental bucket signals $\bm{I}$, parameters $\{\bm\theta,\bm\nu\}$, physical forward model $\bm{H}$, learning rate $\alpha$.
   \State Output: Reconstructed image $\hat{\mathcal{Y}}$.
   \State Parameters initialization: $\bm\theta,\bm\nu \leftarrow \textsc{random(0,1)}$;
\State $ \mathcal{L}\leftarrow \infty$, $t\leftarrow 0 $;
\While{$\mathcal{L}^{(t)} > \varepsilon$ \text{and} $\left\|\partial_{\bm{\theta},\bm{\nu}} \mathcal{L}^{(t)}\right\|^2 > \epsilon$ \text{and} $t \leq T$  
}
\State Calculate $\hat{\mathcal{Y}}^{(t)}$ with Eq. (\ref{CNN-output}) by using $\langle \mathcal{O}\rangle_{\bm z, \bm\theta}^{(t)}$;
\State $\mathcal{L}^{(t)} \leftarrow \left\|\bm{I}-\bm{I}'_t\right\|_2^2$;
\State Evaluate $\partial_{\bm{\nu}}\mathcal{L}^{(t)}$ using Eq.~(\ref{CG});
\State Evaluate $\partial_{\bm{\theta}}\mathcal{L}^{(t)}$ using Eq.~(\ref{QAD})-Eq.(\ref{PSR-f});
\State $\bm{\theta}^{(t+1)} \leftarrow \bm{\theta}^{(t)} - \alpha \partial_{\bm\theta} \mathcal{L}^{(t)}(\bm\theta,\bm\nu)$;
\State $\bm{\nu}^{(t+1)}  \leftarrow \bm{\nu}^{(t)} - \alpha \partial_{\bm\nu} \mathcal{L}^{(t)}(\bm\theta,\bm\nu)$;
\State 	$t\leftarrow t+1$;
\EndWhile
   \end{algorithmic}
\end{algorithm}

\section{Results}\label{R-D}

\subsection{Experimental system}

\begin{figure*}[htp]
	\centering
	\includegraphics[width=0.7\textwidth]{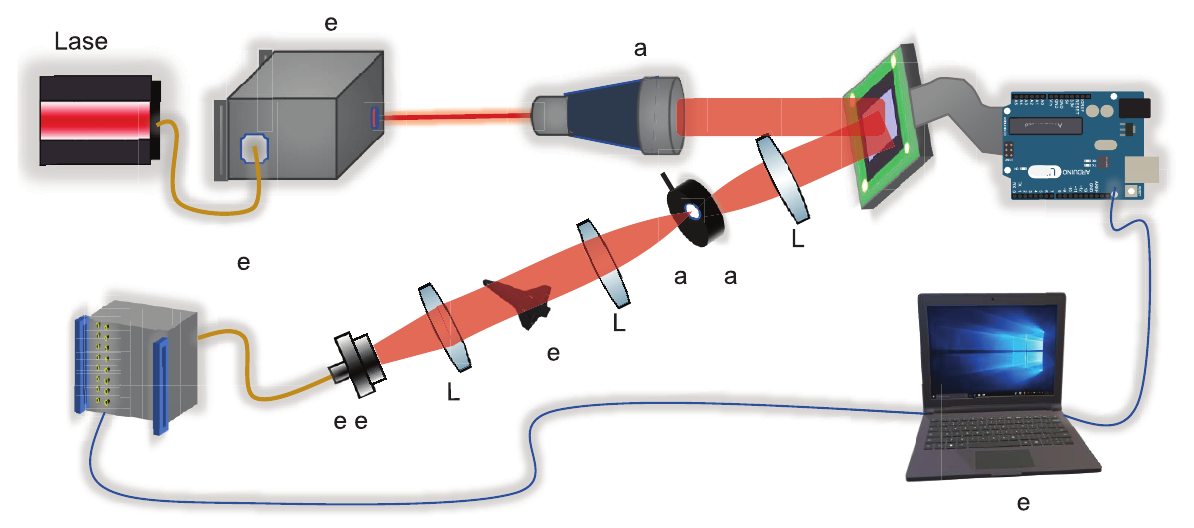}
	\caption{Experimental setup of computational ghost imaging system with a programmable DMD.}
	\label{fig:2}
\end{figure*}

We build a conventional single-beam computational ghost imaging system for data acquisition. 
As shown in Fig. \ref{fig:2}, the beam from a pulsed laser (SC-OEM, YSL) first passes through the fiber into the filter (AOTF, YSL) and then is filtered with the wave length $\lambda = 680$ nm. Next, the monochromatic light beam expanded by a collimator for illuminating the DMD (DLC9500P24, 1080$\times$1920). The DMD is loaded with a sequence of patterns with the number of $M$. After that, the reflected light of DMD is modulated spatially and propagates to the object by a 4f system (L1 and L2) and a diaphragm. After interacting with the object, the light is collected by L3 and detected by a single photon avalanche diode (SPAD) and finally feed into the photon counter (HydraHarp 400) to obtain the bucket signals. The focal lengths of  L1, L2 and L3 are all 100 mm. After data acquisition, the bucket signals with the number of $M$ and the corresponding illumination patterns are used for image reconstruction.

\begin{figure*}[htbp]
	\centering
	\includegraphics[width=1.0\textwidth]{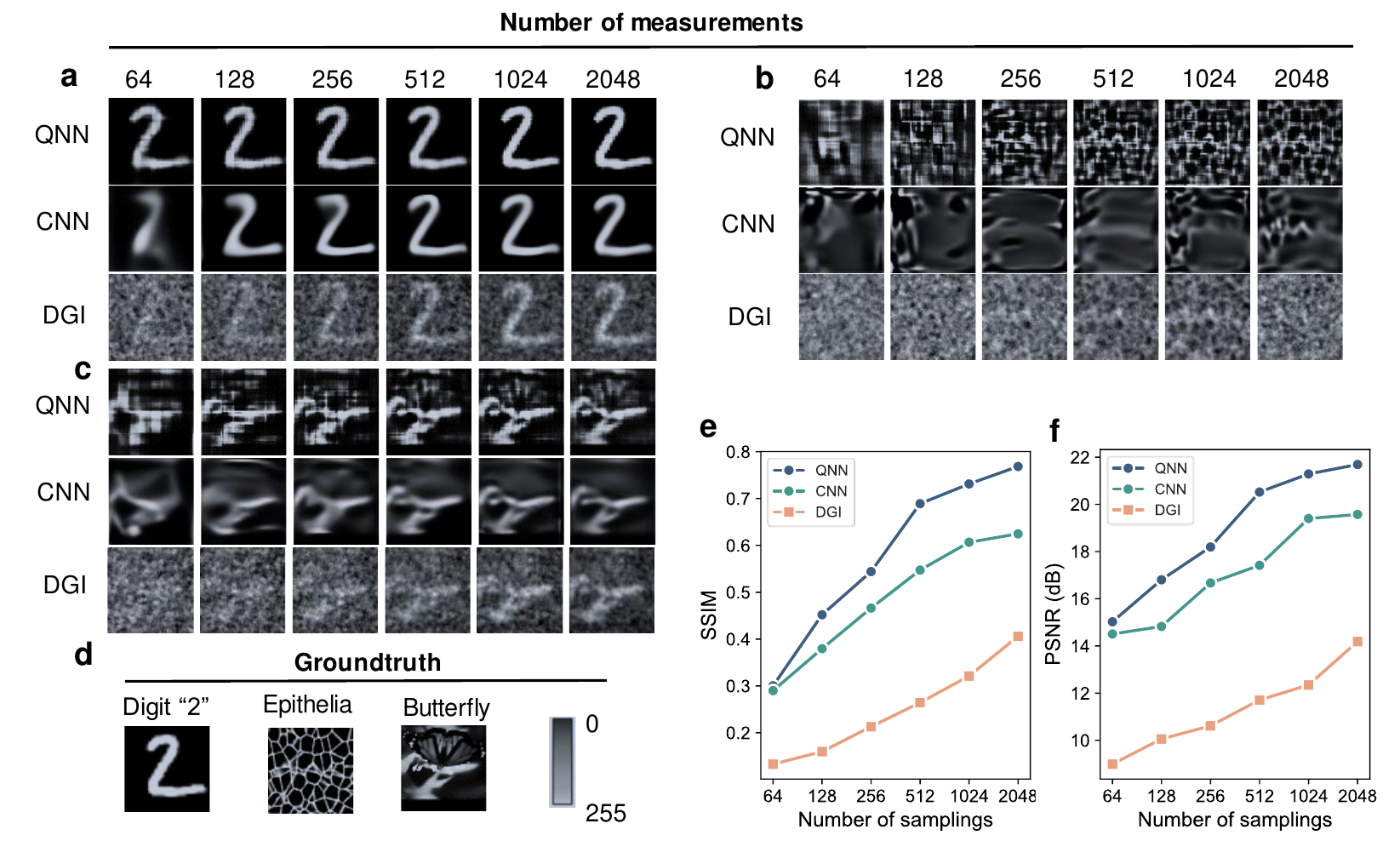}
	\caption{Experimental results of DGI, CNN, QNN  under the different number of  samplings. (a), (b), and (c) are imaging results of digit "2", Epithelia, and Butterfly, respectively. (d) The ground truths of imaging objects. (e) The SSIM index of reconstructing "Butterfly" at different sampling numbers. (f) The PSNR value of reconstructing "Butterfly" at different sampling numbers. The image size is $64 \times 64$ and the iteration number is 1000.}
	\label{fig:3}
\end{figure*}

\subsection{Results under different number of measurements}
In this section, we demonstrate the reconstruction results of different objects using various methods at different number of measurements. We compare the performance of three reconstruction schemes: QNN, CNN,  and DGI\cite{ferri2010differential}. We note that the results of QNN are obtained by using angle-encoding strategy with re-uploading style. In Appendix \ref{more_exp}, we also conduct experiments under physical-inspired encoding strategies to analyze the reconstruction performance, which shows the similar results with angle encoding. For clarity, we do not discuss the results in the main text. The results are shown in Fig. \ref{fig:3}. From Fig. \ref{fig:3}d, it can be found that the reconstruction quality of all methods gradually converges to the groundtruth of the objects as the number of samples increases. Both QNN and CNN achieve clean results, while DGI is limited by compressed sampling making the reconstruction results noisy. Since reconstruction quality is related to target complexity (e.g., sparsity, gray scale), here we compare three types of targets: a simple binary target (digit "2"), a complex binary target (Epithelia image), and a grayscale target (Butterfly image). The first group of the results is plotted in Fig. \ref{fig:3}a.  One can clearly see that the simple binary object has been successfully reconstructed at all sampling numbers by QNN even at the number of measurements as low as 64 at which the compressing ratio is $M/N = 1.56\%$. The results of CNN  are not good compared to the results of QNN when the number of samples are 64  and 128. The second group of the results is plotted in Fig. \ref{fig:3}b. Similar to the results of the first group, the results of QNN are closer to the groundtruth than those of CNN at the same number of samplings. And since the binary target in this group is more complex, the good results can be reconstructed at a higher sampling numbers by QNN, e.g., $M=256$. The last group of the results is plotted in Fig. \ref{fig:3}c. One can observe the same results in the cases that the object is complex binary. For most cases of number of measurements, QNN outperforms CNN and DGI both in terms of visual appearance.

We also quantitatively analyzed the imaging quality of the different methods at different sample numbers. As shown in Fig. \ref{fig:3}e and f, we use quantitative evaluation indexes, the peak signal-to-noise rate (PSNR) and similar structure (SSIM) \cite{hore2010image}. The results show that the PSNR and SSIM of QNN and CNN imaging results for butterfly are significantly higher than those of DGI. And QNN is consistently leading in both PSNR and SSIM compared to CNN. In addition, we can see that the gap in imaging quality between QNN and CNN is getting bigger as the number of samples increases. 
In addition, as a potential algorithm for the application of quantum neural network, the proposed QCSGI will perform more faster with real quantum resources. 

\begin{figure*}[htbp]
	\centering
	\includegraphics[width=0.9\textwidth]{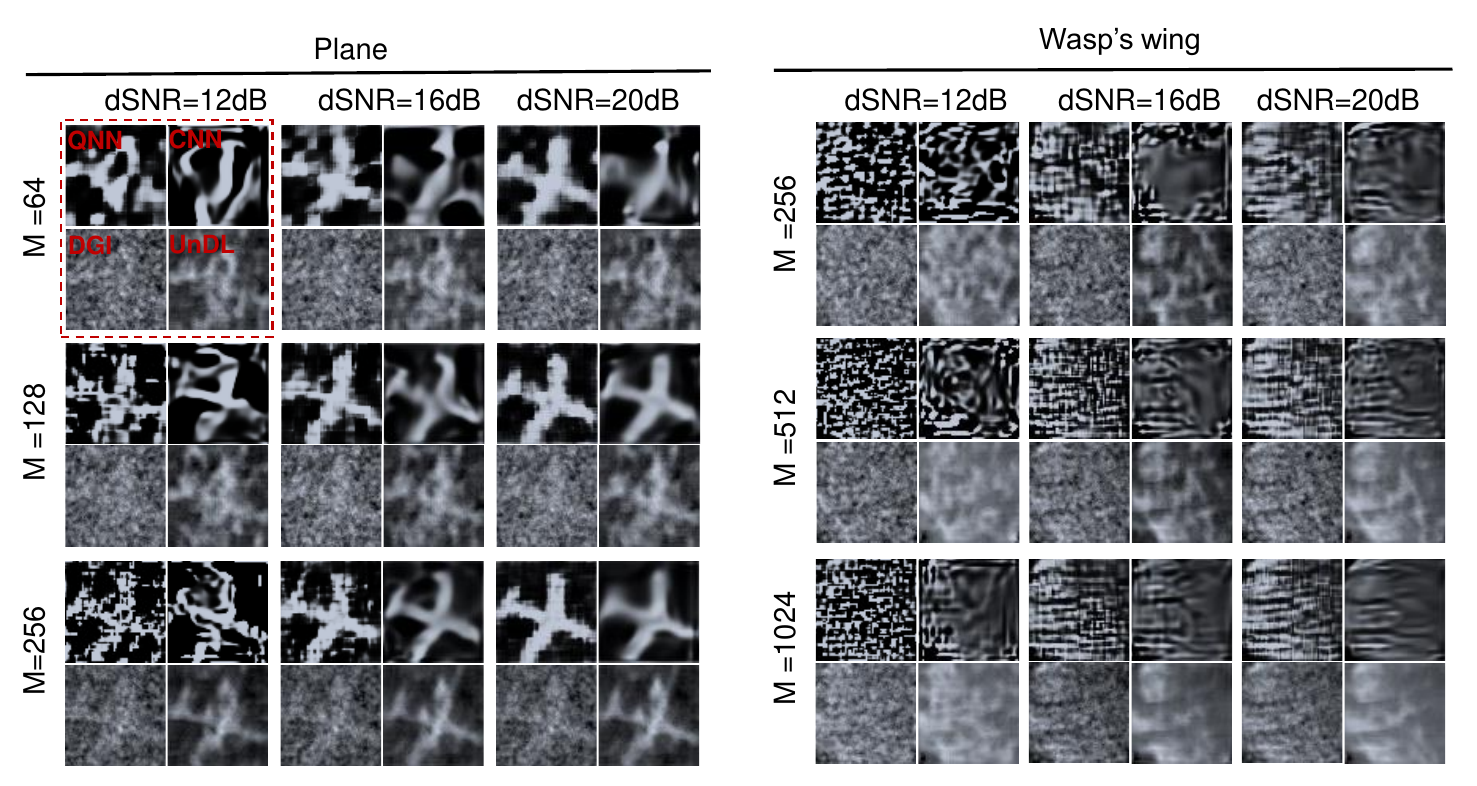}
	\caption{Experimental results of QNN, CNN, DGI, and UnDL under the different numbers of  samplings and different dSNRs of two objects: a plane and a wasp's wing. Under the same dSNR setting and the same number of samples, the four results are reconstructed by QNN (left top), CNN (right top), DGI (left bottom), and UnDL (right bottom). The image size is $64 \times 64$ and the iteration number is 1000.}
	\label{fig:4}
\end{figure*}
\subsection{Anti-noise performance analysis}
To evaluate the robustness of the proposed method, here we conduct the imaging experiments under different noise levels and compare the proposed QCSGI with DGI \cite{ferri2010differential}, a CNN-based model and a conventional unsupervised model. Due to the complexity of noise sources in imaging systems, it is difficult to precisely control the noise level in experiments. Therefore, we model the noise effect as an additive Gaussian distribution to explore the imaging capability under different noise levels. Specifically, since detection noise (shot noise) and background noise exist in the actual imaging environment, we aggregate all these noise and object signals into an intensity sequence. We rewrite Eq. \eqref{eq.1_} and the noise bucket signal can be expressed as 
\begin{equation}\label{eq.1_}
	I_j = \iint H_j(x, y)T(x, y)dxdy + n_j(\sigma),
\end{equation}
where $n_j(\sigma)$ is the additive noise sampled from a zero-mean Gaussian distribution with the standard deviation of $\sigma$. To quantify the noise level, we also define the detection signal-to-noise ratio (dSNR) as
\begin{equation}
	\label{eq.21}
	\text{dSNR} = 10\lg\frac{\langle I_j \rangle}{\sigma}.
\end{equation} 
As can be seen from the Eq. \eqref{eq.21}, the smaller the value of dSNR, the higher the noise level and theoretically the harder it is to reconstruct the image.

Figure \ref{fig:4} shows the imaging results of two object: a simple plane (left) and a wasp's wing (right). We conduct the experiments at four noisy conditions: dSNR $\in \{12, 16, 20 \}$. For reconstructing the plane object, we show the results of sampling numbers of 64, 128, and 256. And for reconstructing the wasp's wing, more samplings are needed because of much detailed information in the target, so we show the results of sampling numbers of 128, 256, and 512. Besides, we compare our proposed method with three typical algorithms. As shown in  Fig. \ref{fig:4}, with the same dSNR setting and the same number of samples, the four results are reconstructed by QCSGI (left top), CNN-based model (right top), DGI \cite{ferri2010differential} (left bottom), and UnDL \cite{zhai2022anti} (right bottom). Noted that UnDL \cite{zhai2022anti} is an unsupervised learning GI method proposed in our former work. For imaging the plane, We can observe that four methods are failed at low dSNR (12dB, 16dB ) with 64 samplings, but QNN can successfully reconstruct the object at dSNR=20 and others three methods still fail. With the increasing of sampling numbers, all methods can still not reconstruct the plane well at dSNR=12. It is intuitive that increasing the number of samples will provide some degree of noise immunity. Therefore,  when dSNR$\leq$ 16, QNN and CNN are able to retrieve  the image with noisy data under the sampling number of 128 and  QNN, CNN, and UnDL can successfully reconstruct the object under the sampling number of 256. For imaging the wasp's wing, the results are similar to imaging airplanes except that more sampling is required.

\begin{figure*}[htbp]
	\centering
	\includegraphics[width=0.9\textwidth]{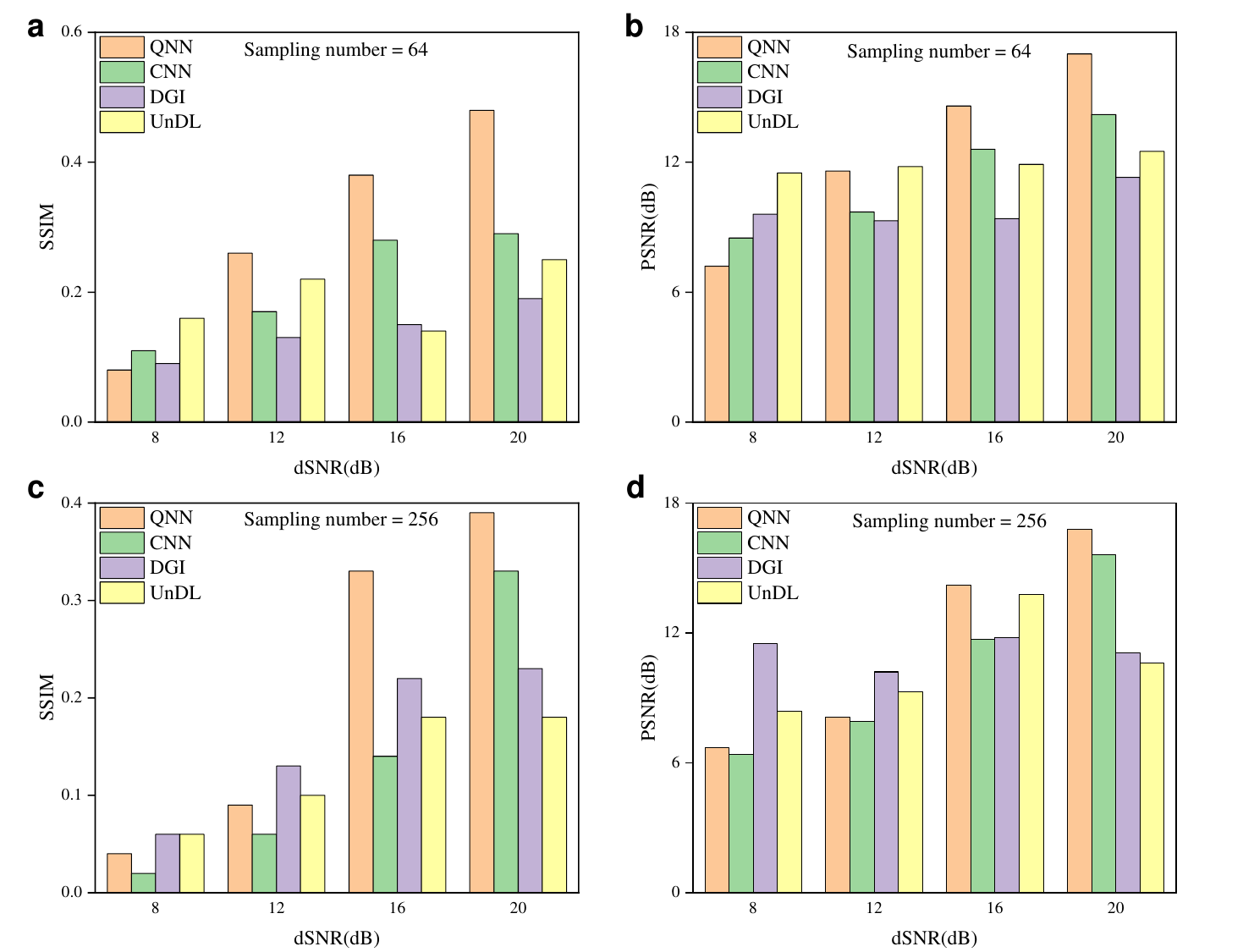}
	\caption{The metrics of different GI reconstruction methods on SSIM and PSNR when
		dSNR is from 8 to 20. (a) and (b) SSIM and PSNR of the object "plane" under $M=64$. (c) and (d) SSIM and PSNR of the object ¡±wasp¡¯s wing" under $M=256$.}
	\label{fig:dsnr}
\end{figure*}

Besides, Fig. \ref{fig:4}  show that CNN results are smoother than QNN's and that QNN's results can show more details, which is consistent with previous results. At this point, we use PSNR and SSIM to quantitatively compare the results, although these two objective image quality evaluation metrics tend to achieve higher scores on smoothed reconstructions. The results is shown in Fig. \ref{fig:dsnr}. One can see that the results of QNN are optimal in most cases, especially when the dSNR is greater than or equal to 12. More results under different numbers of samplings can be found in Appendix \ref{more_exp}.

\subsection{Absence of Barren Plateau and resilience to quantum noise}
\begin{figure*}[htbp]
	\centering
	\includegraphics[width=0.95\textwidth]{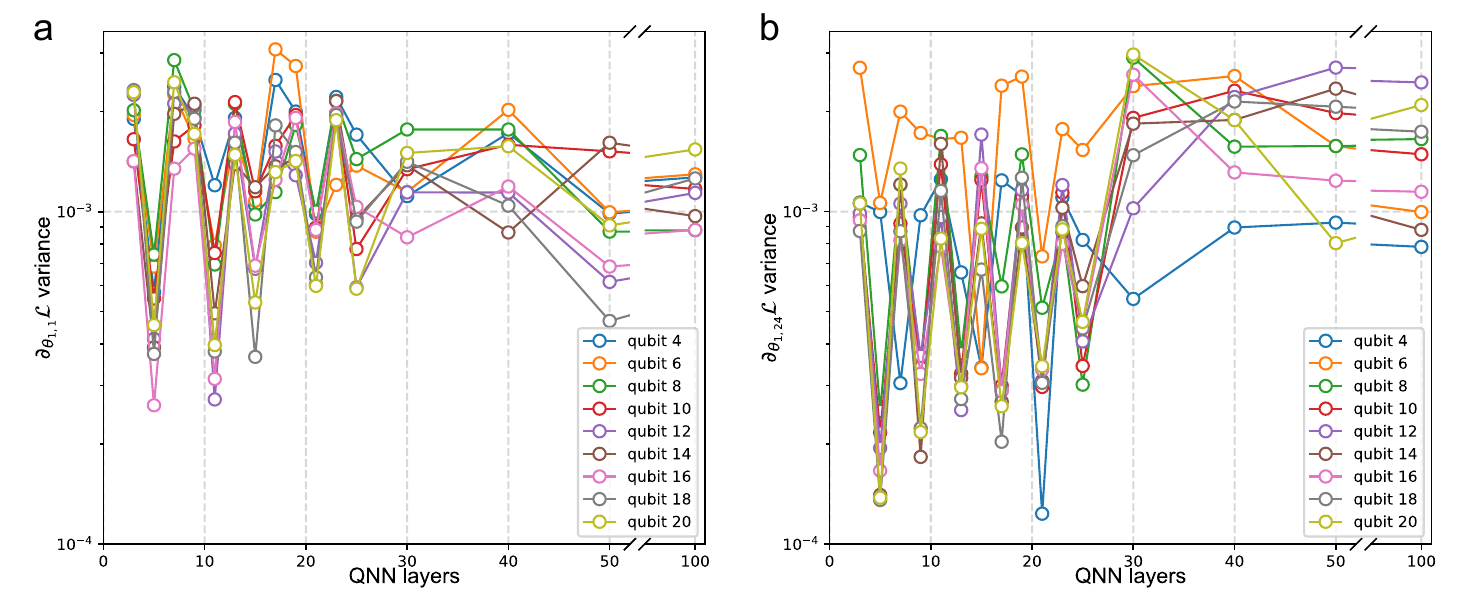}
	\caption{The variance of gradients over local and entanglement parameters of physical-enhanced QNN for random initialized quantum circuits. (a) The gradient of the physical enhanced loss function over the local parameter $\theta_{1,1}$ (i.e. the first local parameter in the first layer). (b) The gradient of the loss function over the entanglement parameter $\theta_{1,24}$ (i.e. the first entanglement parameter in the first layer). The gradient variances for each qubit and layer are evaluated with 100 random quantum circuits.  }
	\label{fig:BP_HQNN}
\end{figure*}

In general, random initialized quantum circuits readily suffer from the barren plateau (BP) problem. More formally, the gradient of variational parameters decreases exponentially as the number of qubits and layers increase. This problem hinders the ability of QNN in ground energy solving and other machine learning approaches. BP is caused by the high expressive power of QNNs such that the loss landscape is extremely flat in large qubits or deep layers. Moreover, quantum circuit noise can also induce the BP in variational quantum algorithms. Using local cost function i.e. projecting the large feature space into local space is proven to be useful to mitigate the problem \cite{cerezo2021cost,ragone2024lie}. In addition, one may also conduct appropriate parameter initialization to mitigate the problem \cite{liu2023mitigating}. Here, we make use of the physical enhanced loss function, which is different from the loss function of Hamiltonian expectation in VQE. Moreover, the feature map of the QNN is further reshaped by the classical neural network, which is likely to be beneficial for escaping from the local minima. The gradient variance of our proposed algorithm for local and entanglement parameters are shown in Fig. \ref{fig:BP_HQNN}. In  Fig. \ref{fig:BP_HQNN}(a), the gradient variances for different qubits and layers are oscillating around $10^{-3}$. When we increase the layers to 100, the gradient variance for large qubits such as 20, is not decreased to $10^{-4}$ or even smaller. In  Fig. \ref{fig:BP_HQNN}(b), we show the gradient variance of entanglement parameters. In general, the entanglement parameter is more sensitive to the loss function compared to the parameters of single-qubit gate. The simulation results validates this expectation. Under the same simulation setting with Fig. \ref{fig:BP_HQNN}(a), the gradient variance also oscillates around $10^{-3}$ but with a larger amplitude. It is noteworthy that the gradient variances of small qubit numbers ($<16$) uses very few measurement samples. Under this setting, the image cannot be recovered. However, the imaging ability should be distinguished from the BP problem. Our main point is to emphasize that the physics-enhanced loss function rather than the variational cost function (such as VQE) is also able to mitigate or even eliminate the BP problem. We use a patch strategy to process the high dimensional bucket signals and each patch owns $16$ qubits. In principle, the gradient variance of patched quantum circuits owns the same scaling with the circuits that have the same number of qubits. \textcolor{black}{The proposed physics-enhanced loss function can be generalized to other inverse problems where the forward process model is known and the inverse process needs to be solved. Such problems often involve working with underdetermined matrices. Reparameterizing the inverse process using QNNs may offer new opportunities for solving this challenge. Additionally, for these problems, we speculate that constructing a loss function based on the physical forward model can be used to optimize the QNN, helping to alleviate the BP issue to some extent. This approach allows for better utilization of quantum computing capabilities.} 

The impact analysis of quantum noise on QCSGI is demonstrated in Appendix \ref{more_exp}. The results imply that the reconstruction performance of QNN-CSGI is not sensitive to quantum noise.

\section{Conclusion}\label{Summary}
In summary, we propose the quantum neural compressive sensing ghost imaging (QCSGI) algorithm to alleviate the stringent requirements of high-quality GI, including large samplings and datasets. Optical GI experiments are conducted across a diverse range of objects, spanning sparse handwritten numbers to complex biological samples. To comprehensively evaluate the performance, we demonstrate the results at a wide range of sample numbers $(64\sim 1024)$. In comparison to conventional methods like differential ghost imaging and classical machine learning-enabled imaging, our algorithm exhibits superior image reconstruction quality, particularly in extremely low sampling situations. Besides, QCSGI effectively recovers images under dSNR conditions with $12\sim 20$ dB, indicating robustness to experimental noise. 

On the quantum algorithm front, our proposed approach differs from VQE-based quantum machine learning and optimization algorithms. QCSGI leverages a physical forward model as an inductive bias to optimize the QNN ansatz, providing robust supervision for the optimization process. In contrast, VQE-based machine learning algorithms necessitate a large labeled dataset for training and VQE-like algorithms for solving ground state and combinatorial problems operate in a variational manner. Through extensive numerical simulations with experimental data, we observe that QCSGI avoids the Barren Plateau problem, i.e. the gradient over quantum parameters does not exponentially vanish with the linear increase of ansatz size. More importantly, QCSGI can still recover images under different quantum noise levels, a result likely attributed to the absence of the Barren Plateau, rendering it highly suited for current NISQ devices. Additionally, the resilience to quantum noise can reduce the need for quantum error mitigation or correction.

The proposed algorithm finds potential applications in mid-frame and terahertz imaging, offering cost reduction for imaging systems and improved imaging quality. Furthermore, our algorithm contributes to the research fields of physics-driven machine learning and optimization \cite{karniadakis2021physics, yu2022gradient} and computational optical imaging \cite{barbastathis2019use}. In future work, we aim to explore physical-data-driven quantum machine learning algorithms and their applications in GI and other computational imaging problems.
\begin{acknowledgments}
This work was supported by the National Natural Science Foundation of China (No. 62401359), the fund of the State Key Laboratory of Advanced Optical Communication Systems and Networks, the Innovation Program for Quantum Science and Technology (Grant No. 2021ZD0300703), Shanghai Municipal Science and Technology Major Project (2019SHZDZX01) and SJTU-Lenovo Collaboration Project (202407SJTU01-LR019).
\end{acknowledgments}

\begin{appendix}
\section{Quantum Neural Network} \label{app_qnn}
\subsection{Quantum Encoding Strategies}\label{Encoding}
\begin{figure*}[htbp]
    \centering
    \includegraphics[width=0.7\textwidth]{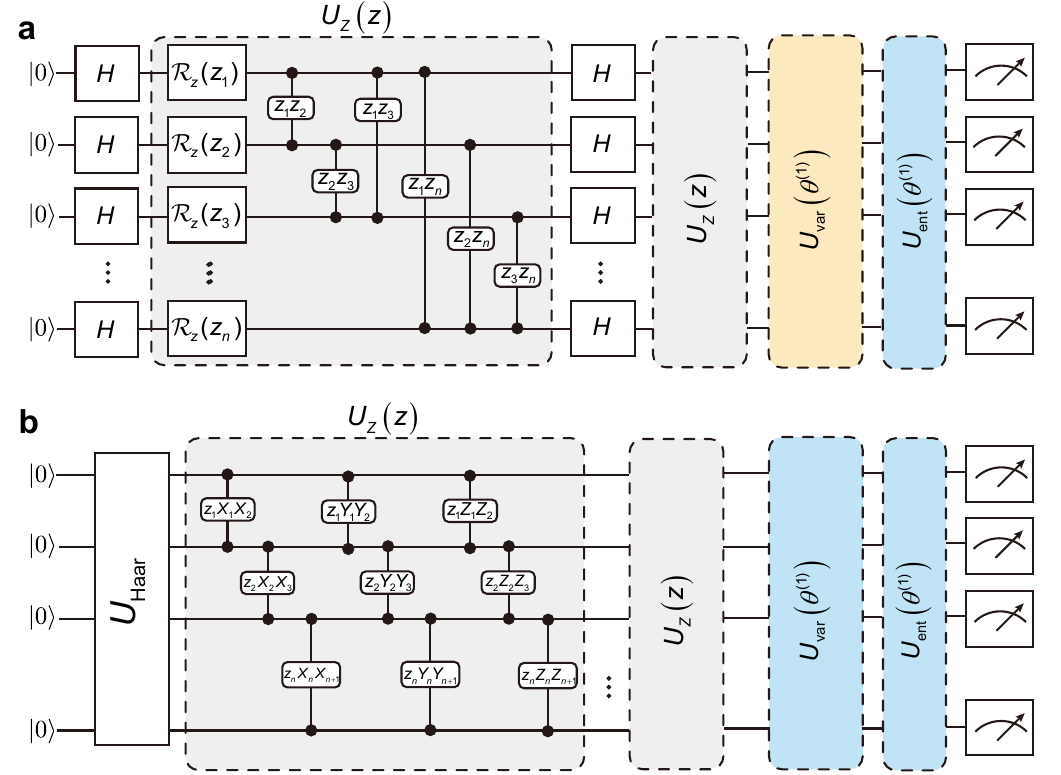}
    \caption{The detailed structure of parameterized quantum circuits. (a) Instantaneous quantum polynomial (IQP) encoding strategy $\mathcal{U}_Z(\bm z)$ requires full connection of qubits. For quantum processors without full connectivity, it is required using SWAP operations to achieve arbitrary two qubit entanglement. (b) Heisenberg-evolution encoding strategy, which only requires neighboring connection to achieve two qubit entanglement operation. $U_{\text{Haar}}$ is the unitary operation for generating the Haar-random quantum state $|\varphi \rangle $.}
    \label{fig:PQC_detail}
\end{figure*}

 Schuld et al. \cite{schuld2021effect} analyze the effect of different quantum encodings on the expressive power of QNNs from the perspective of Fourier spectrum. The frequency spectrum of the learned function is solely determined by the eigenvalues of the data-encoding Hamiltonians. Moreover, the specific structure of the quantum circuit determines the Fourier coefficients. Therefore, the data-encoding and the circuit structure together determine the Fourier transformation of the learning function. Typically, there are three encodings such as the amplitude encoding, angle encoding and instantaneous quantum polynomial (IQP) encoding \cite{havlivcek2019supervised}. These encodings have different encoding Hamiltonians. The merit of the amplitude encoding is the exponential efficiency in encoding the classical data. Suppose the number of features is $\bm{z}\in \mathbb{R}^M$, it is only required $n=\lceil  \log_2 M\rceil $ qubits to encode the classical feature vector. However, the amplitude encoding requires the quantum random access memory, thus is not applicable in the NISQ era. Angle encoding is the most widely used strategy in the NISQ machine learning field. The encoding Hamiltonian of the $i$th term is $H^{(i)}_{enc} = z_iP_i$ where $P_i\in \{I,X,Y,Z\}$. The total Hamiltonian for encoding the full feature vector is $H_{enc} = \otimes_i H^{(i)}_{enc}$. It can be found that the encoding strategy is local with respect to each feature element. This encoding strategy is relatively easier to be implemented in the NISQ machines. IQP encoding is inspired from the 2D many body Hamiltonian. 
 The encoding circuit is shown in Fig. \ref{fig:PQC_detail}b. The encoding and learning operations are separated. The quantum operation of the IQP encoding is given by
\begin{equation}
	|\Psi_{\text{IQP}}\rangle = \mathcal{U}_{Z}(\bm{z}) H^{\otimes n} \mathcal{U}_Z(\bm{z})H^{\otimes n} |0\rangle^{\otimes n},
\end{equation}
where $H^{\otimes n}$ is the unitary that applies the Hadamard gates on all qubits in parallel, and
\begin{equation}
	\mathcal{U}_Z(\bm{z}) = \exp\left(\sum_{j=1}^n z_{i}Z_j +\sum_{i=1}^n\sum_{i'=1}^n z_iz_{i'}Z_iZ_{i'} \right),
\end{equation}
with $Z_i$ representing the Pauli-Z operator acting on the $i$-th qubit. Note that the data range of input $s$ is in $[0,2\pi]^n$ in the original proposal to match the unitary with
\begin{equation}
    \mathcal{U}'_Z(\bm{z}) = \exp\left(\sum_{j=1}^n z_{i}Z_j +\sum_{i=1}^n\sum_{i'=1}^n (\pi-z_i)(\pi-z_{i'})Z_iZ_{i'} \right).
\end{equation}
In our task, we normalize the bucket signals into the range of $(0,1)$. Therefore, we made the equivalent changes to the definition of  $\mathcal{U}_Z(x)$. When the data encoding step is finished, the subsequent quantum learning operations are followed. We make use of the same learning operations as the data reuploading circuit as Fig. \ref{fig:PQC_detail} shows.
 Exactly calculating the IQP encoding in the quantum circuit with large scale qubits on classical computer is \#P hard. In \cite{havlivcek2019supervised}, the researchers have implemented the quantum machine learning experiments in superconducting quantum computer with 2 qubits encoding the classical feature with IQP circuit. 
 
\begin{figure}[htbp]
    \centering
    \includegraphics[width=0.4\textwidth]{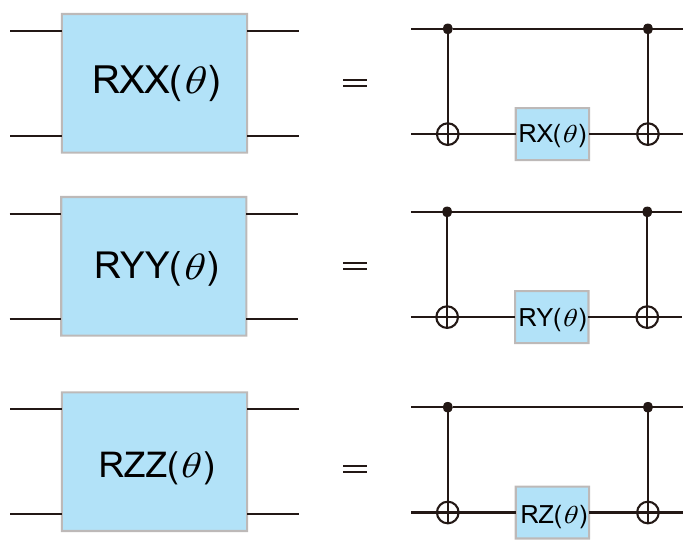}
    \caption{The gate decomposition of RXX, RYY and RZZ gate into three elementary gates.}
    \label{fig:gate_decomp}
\end{figure}

 The Heisenberg-evolution encoding makes use of a Hamiltonian evolution ansatz which is well-studied in the literature of quantum many-body physics \cite{wecker2015progress, cade2020strategies, wiersema2020exploring}. We consider a Trotter formula with $T$ Trotter steps for evolving a 1D-Heisenberg model with interactions given by the classical input $\bm{z}$ for a time $t$ proportional to the system size
\begin{equation}
	\left|\Psi_{\text{HG}}\right\rangle=\left(\prod_{i=1}^n \exp \left(-\mathrm{i} \frac{t}{T} z_{i}h_ih_{i+1}\right)\right)^T \bigotimes_{i=1}^{n+1}\left|\varphi_i\right\rangle
\end{equation}
where $h_ih_{i+1} = X_i X_{i+1}+Y_i Y_{i+1}+Z_i Z_{i+1}$ denotes the neighbor Pauli interactions, $|\varphi_i\rangle$ is a Haar-random single qubit quantum state, which can be prepared with the Haar unitary acting on $|0\rangle$ state. We sample and fix the Haar-random quantum states $|\varphi_i\rangle$ for every qubit. In our simulation, we let $T=3, t=n/3$. The number of qubits when using Heisenberg-evolution encoding requires one extra qubit compared to IQP or angle encoding styles. The following learning operations are the same with the data-reuploading circuit. When implementing the RXX, RYY and RZZ gates, we can use two CNOT gates with one Pauli rotation gate to equivalently achieve the same transformation. Specifically, RZZ gate can be decomposed into three elementary gates,  
\begin{equation}\label{A5}
    \text{RZZ}(\theta) = \text{CNOT} (I\otimes \mathcal{R}_Z(\theta)) \text{CNOT}.
\end{equation}
RXX and RYY gates can also be decomposed into Eq. (\ref{A5}) by replacing the RZ gate with RX and RY gates respectively. The schematic of RXX, RYY and RZZ gates and their decomposition are shown in Fig. \ref{fig:gate_decomp}.

 In Ref. \cite{abbas2021power}, the authors have analyzed the power of the QNN through the quantum Fisher information which is similar to the analyze of Fourier spectrum. Their study reveals that the entanglement encoding is likely to enhance the expressive power since the spectrum of the quantum Fisher information is broaden compared to the local quantum encodings such as the angle encoding. However, we find that in practical scenarios, different encoding strategy have differed performance which is varied with the data set structure \cite{wang2023hybrid}. Moreover, the simulation results in Ref. \cite{xiao2023practical} show that complex entangled encoding strategy such as the IQP encoding and Heisenberg-evolution encoding  \cite{huang2022quantum} is not necessary to show better performance. In contrast, these complex encoding strategy may increase the hardness and indistinguishability of the classical features in quantum Hilbert space. 

\subsection{Expressive power}
We regard the QNN as the parameterized ansatz or constraint in our compressing sensing ghost imaging scheme. By integrating the forward physical operator $H$, we can model the optical imaging process. Therefore, we can calculate the loss function in a unsupervised manner. Compared to quantum machine learning models, i.e. supervised or unsupervised learning, the samples are required to train the gate parameters. In our work, the samples are not required since we can train the QNN with a optimization process. Therefore, we mainly consider the expressive power of the QNN and analyze its potential advantage in solving the inverse problem. Our QNN does not involve the generalization ability of quantum machine learning. The expressive power can be analyzed through the lens of the Fourier spectrum \cite{schuld2021effect}. Generally, the QNN learns a function with the trainable parameter, i.e. $f_\theta(x)=\langle 0|\mathcal{U}^\dagger (x,\theta) O \mathcal{U}(x,\theta)|0\rangle$, where $\mathcal{U}(x,\theta)$ denotes the quantum circuit that depends on the data and a set of parameters. $O$ is some observable, which is generally set to be the local Pauli-z operator. The prediction of the QNN requires repeatably running the quantum circuit multiple times over the measurement results. The quantum circuit generally is composed of multi-layer interleaved encoding and variational layers. Suppose we consider the $m$-subsystem Hamiltonian family $\{H_m|m\in \mathbb{N}\}$, for example the tensor product of Pauli rotations, the Hamiltonian is given by
\begin{equation}
    H^m_{enc} = \sum_{i=1}^m P_i.
\end{equation}
The Fourier spectrum of encoding Hamiltonian $H^m_{enc}$ with eigenvalues $\{\lambda_1, \cdots, \lambda_{d^m}\}$ is given by
\begin{equation}
    \Lambda_{H^m_{enc}} = \{\lambda_j-\lambda_k|j,k\in\{1,\cdots,d^m\}\}
\end{equation}.
For different encoding Hamiltonian, the model class $\{f_m\}$ that the QNN can represent is given by
\begin{equation}
    f_m (x) = \langle 0|U^\dagger_{\text{var}}(\theta) U^\dagger_{ H^m_{enc} }(x) O U_{ H^m_{enc} }(x)U_{\text{var}}(\theta) |0\rangle.
\end{equation}
Intuitively, suppose the learned function has the Fourier spectrum $\mathbb{Z}_K = \{0,\pm K|K\in \mathbb{N}\}$, hence as long as the Fourier spectrum of the Hamiltonian familiar can cover the the spectrum of the learned function, it Hamiltonian family is universal. It can be found that the local Pauli Hamiltonian family is universal when $m=K$, The number of Fourier frequencies grows exponentially as the size of the local subsystems. Therefore, by using a local angle encoding strategy, a variational layer $U_{\text{var}}$ and a observable $O$, the learned function $f_{m'}$ can approximate any function $g\in L_2([0,2\pi]^N)$ with a small error \cite{goto2021universal}. Manzano et al. extends the universal approximating ability into any continuous functions rather than $L_2$ integrable functions \cite{manzano2023parametrized}.

When considering the IQP and Heisenberg-evolution encoding strategy, the Hamiltonian family is more complex than the local Pauli family. The Fourier spectrum thus can cover a wider spectrum.
Therefore, the expressive power of the QNN is universal which can predict any functions with appropriate quantum circuit and encoding strategy.

\subsection{Potential advantage analysis}\label{PA}
From the respective of Fourier spectrum, QNN is capable of universal approximating the functions. GI with low sampling rate requires solving the under-determined equation to obtain the original image. We use QNN to reparameterize the inverse problem. Exactly finding the inverse mapping is hard. However, through mapping the data into the high-dimensional space and optimizing the ansatz, we can reconstruct the image with a low sampling rate. Conventional methods use compressed sensing and convex optimization to recover the original image. Based on the total variation constraint, conventional method can only restore spare images and the sampling rate requirement is also relatively high. By using QNN, a neural network with exponentially large feature space, it is highly likely to better characterize the inverse process. This is one potential advantage source, which is also the potential advantage of numerous quantum machine learning proposals \cite{liu2021rigorous, lee2019experimental} Additionally, the natural representation of QNN as Fourier series suggest that QNN is more suitable for time-series modeling and signal processing \cite{schuld2021effect}. In our GI system, the bucket signals are detected according to different patterns. The hidden structure of the patterns and the bucket signals are more likely to be learned well by QNN. By combining the classical neural network as the post-processing, the hybrid model are more powerful in terms of the expressiveness. Classical neural network can reshape the loss landscape and render it easily to be optimized as the pure QNN may suffer from  Barren Plateau where the gradient vanishes exponentially as the circuit qubit increases. Our hybrid quantum machine learning proposal combining with the physical forward model provides a new path toward to solving the inverse problems.

\section{Quantum noise model}\label{Quconvolution}

In the NISQ era, the effect of quantum noise is the most obstacle for achieving quantum advantage. To study the impact of the quantum noise on the optimization of compressive GI, we simulate the quantum noise in the PQCs by inserting the Markovian quantum noise operator (Kraus operator) between the ideal quantum unitaries. There are two methods to simulate the quantum noise. The first one is to apply density matrix simulation to simulate the full amplitude of quantum noise effect. This method costs $O(4^n)$ with $n$ denoting the number of qubits. Therefore, the simulation is not scalable
on the classical computer. The second method is to use the Monte Carlo simulation in the state vector simulation. This method also will increase the overhead but the overhead is well controlled and acceptable for classical computation. Generally, when the quantum noise is large, any quantum advantage including the computational advantage will disappear since the quantum circuit can be classical efficiently simulated.

When considering the depolarizing quantum channel in the quantum circuit, the Kraus operator is given by 
\begin{equation} \label{QN}
    \Lambda(\rho;\lambda) = \sum_{i=0}^{3} \mathcal{K}_i \rho \mathcal{K}_i^\dagger,
\end{equation}
where $\lambda$ is the depolarizing rate, $\rho$ denotes the original quantum state and the Kraus operators can be given by
\begin{equation}
\begin{split}
     \mathcal{K}_0 = \sqrt{1-\frac{3\lambda}{4}}I,
     \mathcal{K}_1 = \sqrt{\frac{\lambda}{4}} X,
     \mathcal{K}_2 = \sqrt{\frac{\lambda}{4}} Y,
     \mathcal{K}_3 = \sqrt{\frac{\lambda}{4}} Z.
\end{split}
\end{equation}
The Kraus operators satisfy the complete relation i.e. $\sum_i \mathcal{K}_i\mathcal{K}_1^\dagger = I$. Geometrically speaking, the depolarizing channel $\Lambda(\lambda)$ can be interpreted as a uniform contraction of the Bloch sphere, parameterized by $\lambda$. In the case where $\lambda=1$, the channel returns the maximally-mixed state for any input state $\rho$, which corresponds to the complete contraction of the Bloch-sphere down to the single point $\frac{I}{2}$ given by the origin.
The depolarizing rate can be set to different values to simulate different quantum noise level. In addition, we only consider single quantum noise source in the quantum circuit. Since in practical quantum computer, the two qubit gate such as the CNOT gate is easier influenced by the quantum noise. Therefore, we place the depolarization operator after each the entanglement gate. Therefore, the quantum noise is accumulated which can truly reflect the practical situation.

Phase damping and amplitude damping channel are also two methods to characterize the quantum noise sources.
For the amplitude damping channel, it can be written as \begin{equation}
\mathcal{N}(p) = \mathcal{K}_0 \rho \mathcal{K}_0^\dagger + \mathcal{K}_1 \rho \mathcal{K}_1^\dagger, 
\end{equation}
with 
\begin{equation}
    \mathcal{K}_0 = \begin{pmatrix}
    1 & 0  \\
    0 & \sqrt{1-p}
    \end{pmatrix}, \, \mathcal{K}_1 = \begin{pmatrix}
    0 & \sqrt{p} \\
    0 & 0
    \end{pmatrix},
\end{equation}
where $p$ is the amplitude damping parameter which denotes the probability of losing a photo.
For phase damping channel, it can be written as 
\begin{equation}
     \mathcal{K}_0 = \begin{pmatrix}
    1 & 0  \\
    0 & \sqrt{1-p}
    \end{pmatrix}, \, \mathcal{K}_1 = \begin{pmatrix}
    0 & 0  \\
    0 & \sqrt{p}
    \end{pmatrix}.
\end{equation}
It can be found that the phase damping channel does not loss energy into the environment.

In principle, the quantum noise is random for each sample input,
therefore the random seed should be randomly generated to simulate each circuit execution. Overall, the quantum noise will
be canceled out to some extent. This statistical effect comes from the machine learning method, which is also highly similar to
the method of probability error cancellation for error mitigation. Although some studies show that the loss landscape
is exponentially flattened when the quantum noise rate increases in PQC, we deliberate that it depends on the specific problem
and data structure. Integrating classical NN in PQC, i.e. the hybrid quantum-classical neural network is a highly effective way
to overcome the shortcomings of small input size and low-depth circuits while maintaining the superiority of QNN in handling
complex problems such as extracting time series features.

\section{QNN and Optimization}\label{optimization}

In the NISQ era, we usually regard the parameterized quantum circuit (PQC) as QNN. PQC uses the basic quantum gates such as single-qubit rotations and two-qubit entangling gates to compose the trainable QNN. Generally, we make use of hardware-efficient quantum circuit as the ansatz of QNN to learn the patterns hidden in the data. In our imaging scheme, we use the untrained method to optimize the QNN combined with the physics-enhanced prior information. In untrained method, large-scale data set is not required. We regard the QNN model as a surrogate of the inverse optimization process which is a common problem in numerous imaging tasks. 

General QNN consists of quantum data-encoding layers and quantum processing layers as Fig. \ref{fig:1}b shows. In reality, to enhance the expressive power of QNN, we adopt the data-reuploading encoding style. Moreover, since the geometrical limitations of current quantum hardware, we only adopt the neighboring entanglement gates to operate the qubits. Therefore, 
the designed QNN is composed of multi-layer interleaved rotation, entanglement and data-encoding layers. Note that the designed QNN is quantum hard-efficient and can be executed in the NISQ devices. Specifically, the formal expression of the QNN is given by
\begin{equation}
	\mathcal{U}(\bm{z},\bm{\theta}) = {U}_{\text{var}}(\theta_L) \prod_{l=1}^{L-1} {U}_{\text{enc}}(\bm{z}){U}_{\text{var}}(\theta_l),
\end{equation}
where $L$ is the number of layers of the QNN, $\bm{z}\in \mathbb{R}^M$ denotes the input data vector, $\bm{\theta}=\{\theta_1,\cdots, \theta_L\}, \bm{\theta}\in [0,2\pi]^{|\bm{\theta}|}$ represents the trainable parameters, $U_\text{enc}, U_{\text{var}}$ are the encoding and variational layers, respectively. The functionality of QNN is to construct the unitary map give by
\begin{equation}\label{QNN_map}
	|\Psi(\bm{z},\bm{\theta})\rangle = \mathcal{U}(\bm{z},\bm{\theta}) |0\rangle^{\otimes n },
\end{equation}
where $n$ denotes the number of qubits, $(n=M)$ in case with angle encoding, $|\Psi(\bm{z},\bm{\theta})\rangle$ represents the mapped (learned) final quantum state. Equipped with the data-reuploading encoding strategy, the encoding of classical information to quantum state is accomplished by 
\begin{equation}
	U_{\text{enc}} (x) = \bigotimes_{i=1}^M \left[ \mathcal{R}_Y(z_i) \mathcal{R}_Z(z_i)\right].
\end{equation}
Another useful encoding is the amplitude encoding, where one can use $n$ qubits to encode a $2^n$ dimensional data vector with $|z\rangle = \sum_{i=1}^{2^n} z_i|i\rangle$. In theory, amplitude encoding can exponentially save the memory compared to the data re-uploading strategy. However, there is no efficient algorithm to implement the amplitude encoding as it generally requires the quantum random access memory \cite{giovannetti2008quantum}. Physical-inspired quantum encoding such as IQP and Heisenberg-evolution encoding are presented at the  Appendix \ref{app_qnn}.  The parameterized quantum circuit is composed of two fundamental quantum layers i.e. the local qubit rotations and two-local entanglement operations, given by
\begin{equation}
\footnotesize
	U_{\text{var}} (\theta_l) = \bigotimes_{i=1}^{M-1} \text{CZ}_{i,i+1} \bigotimes_{i=1}^M \left[\mathcal{R}_X\left(\theta_l^{i,1}\right)\mathcal{R}_Y\left(\theta_l^{i,2}\right)\mathcal{R}_{Z}\left(\theta_l^{i,3}\right) \right],
\end{equation}
where $\theta_l^{i}$ is composed of three trainable parameters. In reality, we can also make use of CNOT gate or exponential XX or ZZ gate as the entanglement operator. The entanglement geometrical topology can be linear, circular and full style. In the NISQ device, implementing full style entanglement, one requires quantum SWAP operation to realize the non-local entanglement operations.

When the quantum state is learned, we require measuring the quantum state to obtain the classical information. Suppose we use the observable $\mathcal{O}$ as the features. Arbitrary Hermitian operator $\mathcal{O}$ can be decomposed into $\mathcal{O} = \sum_i w_i h_i $ where $w_i$ denotes the decomposition weights, $h_i$ denotes the sub-Hamiltonian in the $n$-qubit Pauli group $h_i\in \mathcal{P}_n$. In order to obtain the expectation value of the observable, we use the quantum expectation estimation method \cite{broughton2020tensorflow} to estimate the expectation value given by
\begin{equation}\label{measure}
	\langle \mathcal{O}\rangle_{\bm{z},\bm{\theta}} = \sum_i w_i \langle \Psi(\bm{z},\bm\theta) | h_i| \Psi(\bm{z},\bm\theta )\rangle = \bm{w} \cdot \bm{h}_{\bm{z},\bm\theta},
\end{equation}
where we define the symbol $\bm h_{x,\bm\theta}$ as the vector of the expectation value of each single sub-Hamiltonian. In fact, the parameters $\bm w$ present the weights of each sub-Hamiltonian and are also trainable to optimize the observations. 

In GI system, we firstly normalize the bucket signal into $[0, 2\pi]^M$ and then map the normalized bucket signals into the QNN. Subsequently, the fully-connected neural network (FNN) is followed to map the one dimensional feature vector into a 2D-shaped tensor. Then, resnet-style convolution neural networks are followed to process the reshaped tensor to obtain the final reconstructed image. The process is displayed as Fig. \ref{fig:1}b shows. 
However, large-scale QNNs are hard to implement in reality since the errors and the restrictions of coherence time of qubits \cite{cheng2023noisy, bharti2022noisy}. Therefore, we use a qubit patch strategy with which we are able to process the measurement signals separately. Specifically, we can divide the large 1D vector $x$ into small data vectors $\{x^j\}_{j=1}^m$ and each $x^j$ can be processed by independent QNNs so that the final observations set $\{\langle \mathcal{O} \rangle_{x^j,\bm\theta} \}_{j=1}^m$ is obtained by using Eq. (\ref{QNN_map}) and Eq. (\ref{measure}), where $m$ denotes the number of qubit patches. 

The classical part of the hybrid QNN can be presented by a functional mapping $\mathcal{F}$, which consists of fully-connected networks and classical convolution networks. The estimated image is given by 
\begin{equation}\label{CNN-output}
	\hat{\mathcal{Y}} = \mathcal{F}_{\bm \nu}(\langle \mathcal{O}\rangle_{\bm{z},\bm\theta}),
\end{equation}
where $\nu$ denotes the classical trainable parameters. For simplicity, we denote the 
composite functional of the hybrid QNN as 
\begin{equation}
	\Phi_{\bm{\theta},\bm{\nu}} = \mathcal{F}_{\bm{\nu}} \circ \mathcal{U}_{\bm\theta},
\end{equation}
such that the reconstructed image can be represented as $\hat{\mathcal{Y}} = \Phi_{\bm{\theta},\bm{\nu}}(\bm{z})$.

We use gradient descent method to optimize the parameters of the proposed method. The first part is the gradient over the classical parameters $\bm{\nu}$, which is given by
\begin{equation}\label{CG}
    \frac{\partial\mathcal{L}}{\partial \bm\nu} = \frac{\partial \mathcal{L}}{\partial \Phi_{\bm\theta,\bm\nu}} \cdot \frac{\partial \Phi_{\bm\theta,\bm\nu}}{\partial \mathcal{F}_{\bm\nu}}\cdot \frac{\partial \mathcal{F}_{\bm\nu}}{\partial\bm\nu}.
\end{equation}
We note that the gradient of the loss function over the classical parameters can be calculated by the auto-differentiation (AD), which is highly efficient in classical computer. The gradient of the loss function over quantum parameters $\bm\theta$ thus can be calculated by the chain rule, i.e.
\begin{equation}\label{QAD}
     \frac{\partial\mathcal{L}}{\partial \bm\theta}  =  \frac{\partial \mathcal{L}}{\partial \Phi_{\bm\theta,\bm\nu}} \cdot \frac{\partial \Phi_{\bm\theta,\bm\nu}}{\partial \mathcal{F}_{\bm\nu}}\cdot \frac{\partial\mathcal{F}_{\bm \nu}}{\partial \mathcal{U}_{\bm\theta}} \cdot \frac{\partial\mathcal{U}_{\bm\theta}}{\partial\bm\theta}.
\end{equation}
The first three gradients of the right-hand side of the Eq.~ (\ref{QAD}) can be calculated efficiently by using AD. The last partial derivative however should be estimated by the quantum computer. Since the quantum AD is not available, we use the parameter-shift rule \cite{wierichs2022general, schuld2019evaluating} to calculate the partial derivative $\partial\mathcal{U}_{\bm\theta}/\partial\bm\theta$. 
The gradient is defined as
\begin{equation}\label{PSR}
\begin{split}
   & \partial_{\theta_k} \mathcal{U}(\bm{z},\theta_k) = \sum_i w_i \langle \bm{0}|\partial_{\theta_k} \mathcal{U}^\dagger (\bm{z},\theta_k)h_i \mathcal{U}(\bm{z},\theta_k)|\bm{0}\rangle \\
    & = \sum_i cw_i \left[\langle \bm{0}|\mathcal{M}(\theta_k+s)|\bm{0}\rangle - \langle \bm{0}|\mathcal{M}(\theta_k-s)|\bm{0}\rangle\right],
\end{split}
\end{equation}
where we have defined the notation $\mathcal{M}(\bm{z},h_i,\bm\theta) = \mathcal{U}^\dagger (\bm{z},\bm\theta)h_i \mathcal{U}(\bm{z},\bm\theta)$ for simplicity. We note that the coefficient $c$ and shift $s$ are constant and independent of $\theta_k$. Since the QNN is generated by Pauli operations, suppose the QNN is composed of $\mathcal{U}({\theta_k}) = \exp(-i\theta_k P/2)$. The gradient $\partial_{\theta_k} \mathcal{U}({\theta_k}) = -\frac{i}{2} \mathcal{U}({\theta_k}) P$ and inserting into Eq. (\ref{PSR}), we have 
\begin{equation}
    \partial_{\theta_k} \mathcal{U}(\bm{z},\theta_k) = -\frac{i}{2} \sum_j w_j \langle \bm{0} | \mathcal{U}^\dagger(\theta_k) [P, h_j] \mathcal{U}(\theta_k)|\bm{0}\rangle. 
\end{equation}
With commutation rule, i.e. $[P,h_i] = i\mathcal{U}^\dagger(\frac{\pi}{2})h_i\mathcal{U}(\frac{\pi}{2}) - \mathcal{U}^\dagger(-\frac{\pi}{2})h_i\mathcal{U}(-\frac{\pi}{2}) $, we can derive the analytical gradient as 
\begin{equation}\label{PSR-f}
    \partial_{\theta_k} \mathcal{U} =  \sum_i \frac{w_i}{2} \left[ \langle \bm{0}|\mathcal{M}(\theta^+_k)|\bm{0}\rangle - \langle \bm{0}|\mathcal{M}(\theta_k^-)|\bm{0}\rangle \right],
\end{equation}
 where $\theta_k^\pm = \theta_k \pm \pi/2 $. As the special property of Pauli operations, we are able to provide an analytical evaluation of the gradient of the expectation over quantum parameters. The time complexity of evaluation quantum gradient is determined by the total number of quantum parameters. More detailed about the convergence and complexity analysis can be found in Appendix \ref{convergence}.

\section{Convergence and complexity analysis}\label{convergence}
The Algorithm \ref{algorithm} iterates through line 4-13 until a stopping criterion is met i.e. the gradient norm is smaller than $\epsilon$ or the MSE smaller than $\varepsilon$ or the maximum number of iterations reached. The stopping criterion ensures that the algorithm terminates after a reasonable amount of time, even if the optimal solution has not been found. To demonstrate the convergence of the quantum gradient and classical gradient descent algorithm, we will show the cost function $\mathcal{L}(\bm{\theta},\bm{\nu})$ converges to a minimum value under suitable conditions.
\begin{figure*}[htbp]
    \centering
    \includegraphics[width=1\textwidth]{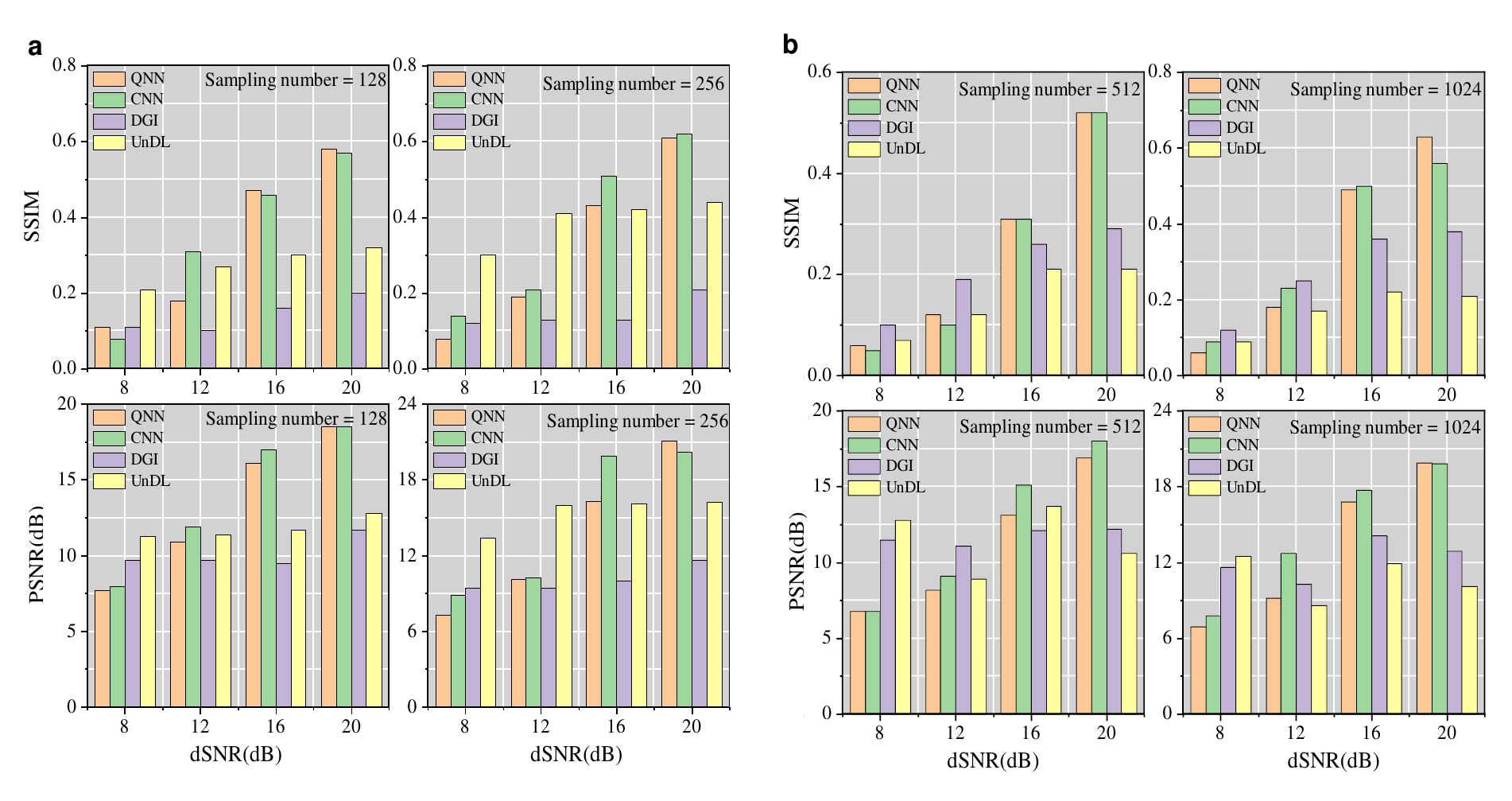}
    \caption{The metrics of different GI reconstruction methods on SSIM and PSNR when
 dSNR is from 8 to 20. (a) SSIM and PSNR of the object ”plane" under $M=128$ and $M=256$. (b) SSIM and PSNR of the object ”wasp’s wing" under $M=512$ and $M=1024$.}
    \label{fig:dsnr_supp}
\end{figure*}

Let $\mathcal{L}:\mathbb{R}^M\rightarrow \mathbb{R}$ be a differential loss function, and suppose the initial classical and quantum parameter set are denoted as $\bm{\xi}^{(0)}=(\bm{\theta}^{(0)},\bm{\nu}^{(0)})$. In case $\mathcal{L}$ is convex and has a Lipschitz continuous gradient with constant $K$, we have
\begin{equation}
    \mathcal{L}\left(\boldsymbol{\xi}^{(t+1)}\right) \leq \mathcal{L}\left(\boldsymbol{\xi}^{(t)}\right)-\frac{\alpha}{2}\left|\nabla \mathcal{L}\left(\boldsymbol{\xi}^{(t)}\right)\right|^2+\frac{K \alpha^2}{2}\left|\nabla \mathcal{L}\left(\boldsymbol{\xi}^{(t)}\right)\right|^2 .
\end{equation}
Rearranging the inequality and summing over all the iterations, we obtain
\begin{equation}
    \sum_{t=0}^{T-1}\left|\nabla \mathcal{L}\left(\boldsymbol{\xi}^{(t)}\right)\right|^2 \leq \frac{\mathcal{L}\left(\boldsymbol{\xi}^{(0)}\right)-\mathcal{L}\left(\boldsymbol{\xi}^\star \right)}{\alpha}+\frac{K \alpha}{2} \sum_{t=0}^{T-1}\left|\nabla \mathcal{L}\left(\boldsymbol{\xi}^{(t)}\right)\right|^2,
\end{equation}
where $\bm{\xi}^\star$ denotes the global minimum of $\mathcal{L}$. Rearranging the inequality and the sum of squared gradients is given by
\begin{equation}
    \sum_{t=0}^{T-1}\left|\nabla \mathcal
    {L}\left(\boldsymbol{\xi}^{(t)}\right)\right|^2 \leq \frac{\mathcal
    {L}\left(\boldsymbol{\xi}^{(0)}\right)-\mathcal
    {L}\left(\boldsymbol{\xi}^\star\right)}{\alpha\left(1-\frac{K\alpha}{2}\right)},
\end{equation}
where $K\alpha < 2$ since $\mathcal{L}$ has a Lipschitz continuous gradient. As the number of iterations $T$ appraoches infinity, the left-hand side of the inequality converges to zero, which implies that the gradient of the cost function $\nabla\mathcal{L}(\bm{\xi})$ converges to zero. Consequently, the quantum gradient descent algorithm converges to a global minimum of the cost function $\mathcal{L}(\bm{\xi})$.

The number of iterations required for convergence depends on the choice of the
learning rate $\alpha$, the Lipschitz constant $K$, and the desired accuracy $\epsilon$. Suppose we have $\mathcal{L}(\bm{\xi}^{(T)}) - \mathcal{L}(\bm{\xi}^\star) \leq \epsilon$. the upper bound on the number of iterations $T$ is given by
\begin{equation}
    T\leq \frac{\mathcal{L}(\bm{\xi}^{(0)})-\mathcal{L}(\bm{\xi}^{\star}) }{\alpha (1-\frac{K\alpha}{2})\epsilon}.
\end{equation}
The upper bound suggests that the number of iterations is inversely proportional to the desired accuracy. When using the parameter-shift rule to evaluate the quantum gradients, the total cost is determined by the number of quantum parameters. Suppose the required number of quantum circuit shots for each parameter shift is $S$, the total number of shots is thus $O(LMS)$ for hardware-efficient quantum circuit. Consequently, the overall complexity is the product of the complexity of gradient evaluation and the number of iterations, which is given by
\begin{equation}\label{complexity}
    O\left( \frac{LMS\left(\mathcal{L}(\bm{\xi}^{(0)})-\mathcal{L}(\bm{\xi}^\star)\right)}{\alpha (1-\frac{K\alpha}{2})\epsilon}  \right),
\end{equation}
where we neglect the constant and multiplier coefficients. In reality, the actual complexity may be lower due to factors such as the structure of quantum circuit, the choice of the learning rate and the efficiency of the quantum circuit implementation.

\begin{figure*}[htbp]
    \centering
    \includegraphics[width=0.8\textwidth]{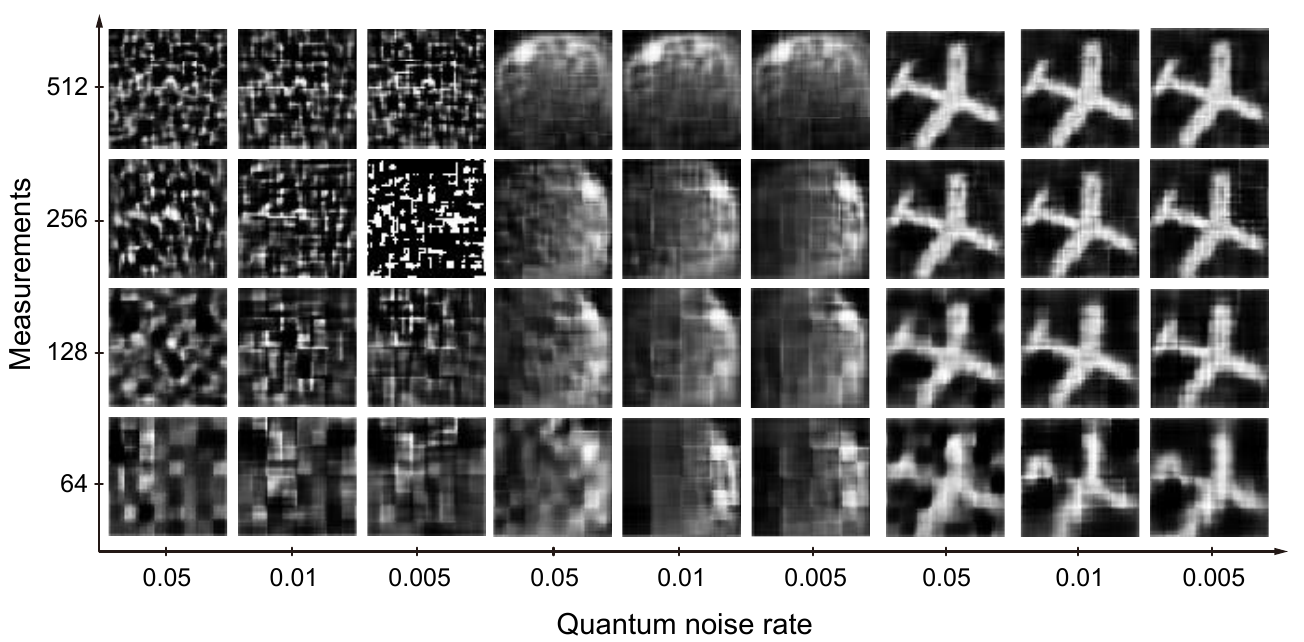}
    \caption{Additional reconstruction results under different quantum noise rates and measurement ratios. Small quantum noise rate overall gives rise to a better reconstruction. Large quantum noise rate also shows a considerable reconstruction performance. }
    \label{fig:Add_noise}
\end{figure*}

\section{More experimental results} \label{more_exp}

\subsection{Anti-noise results under different numbers of samplings}\label{dsnr_supp}
As described in the main text, we use PSNR and SSIM to quantitatively compare the results, although these two objective image quality evaluation metrics tend to achieve higher scores on smoothed reconstructions. Here we show the results of two objects (plane and wasp's wing) under two higher sampling numbers. As shown in \ref{fig:dsnr_supp}, one can see that the results of QNN are optimal in most cases, especially when the dSNR is greater than or equal to 12. in addition, with the increasing of sampling numbers, the performance of CNN has the tendency of catching up to the performance of QNN.

\begin{figure*}[htbp]
	\centering
	\includegraphics[width=0.95\textwidth]{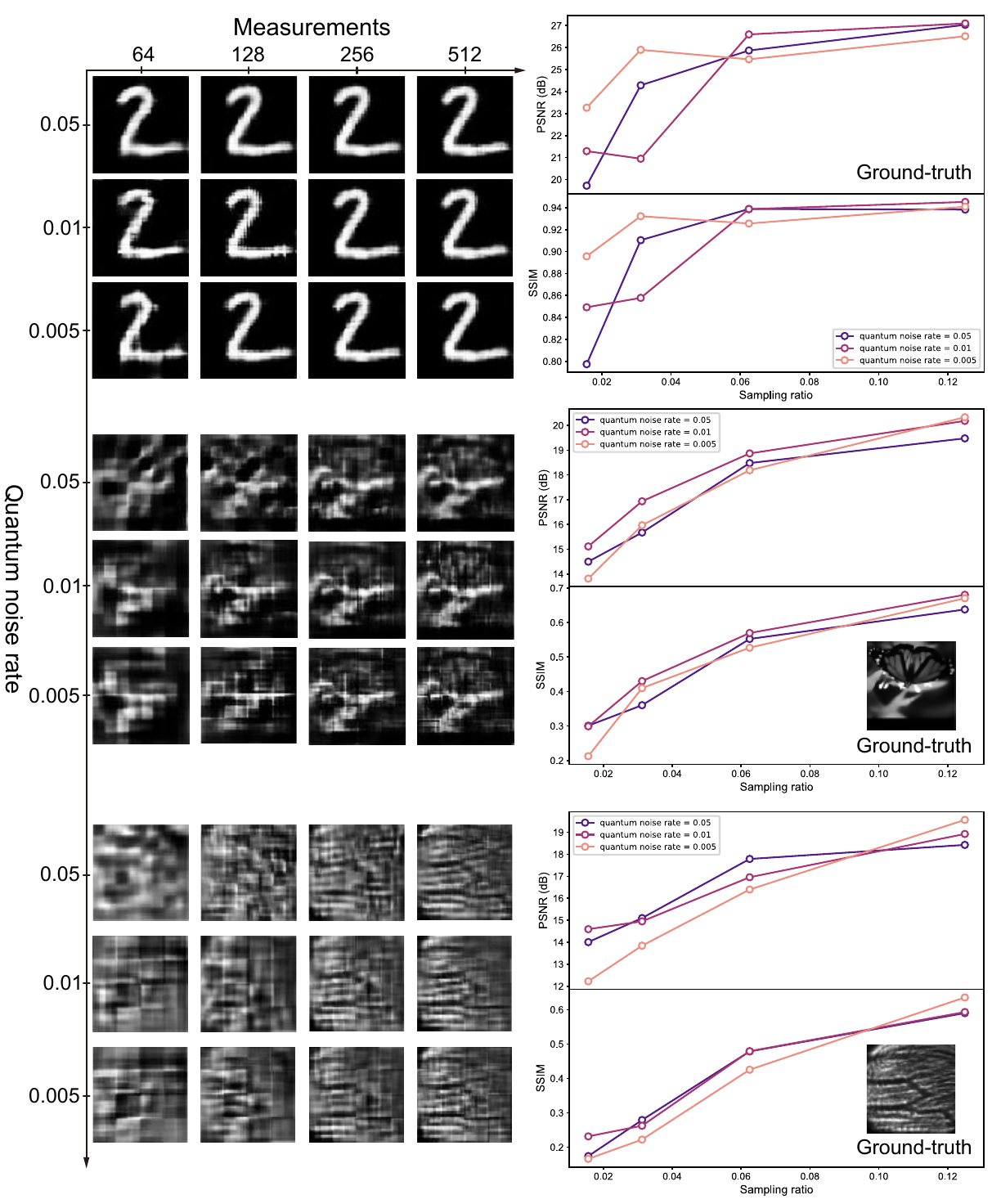}
	\caption{The reconstruction performance of QCSGI (Algorithm \ref{algorithm}) under various quantum noise rates and sample ratios. The peak signal-to-noise ratio (PSNR) and structural similarity (SSIM) metrics are calculated by using the ground-truth image and the reconstructed image. The quantum noise rates are set to be as $\{0.05, 0.01, 0.005\}$ for quantum circuits. The measurements or sampling ratios are set to be as $\{64, 128, 256, 512\}$. The sampling ratio is calculated by \#measurements$/(64\times64)$.}
	\label{fig:noise_qc}
\end{figure*}

\subsection{QCSGI under quantum noisy circuit}\label{rquantumnoise}
In the NISQ era, quantum devices are noisy and the fidelity of quantum gates are not perfect. For single qubit gate, the state-of-art quantum fidelity is $99.99\%$ which can be realized in ion-trap quantum computer. For two-qubit gate, the state-of-art quantum fidelity is $99.9\%$\cite{cheng2023noisy}. In superconducting quantum computer, the fidelity is lower. With the increase of the number of qubits, the fidelity of the whole quantum circuit will decrease dramatically. Optimal strategies to eliminate the quantum noise to obtain the perfect gate fidelity is to use quantum error correction (QEC). However, the overhead of QEC in current quantum devices is prohibitively forbidden for medium-scale quantum devices (>20). Commonly, instead of using QEC, one can use quantum error mitigation (QEM) methods to mitigate the errors. The essence of QEM is to learn the statistical distributions of quantum noise and uses the learned function to filter (correct) the experimentally estimated observables. Nevertheless, the overhead is still exponentially grown for the NISQ devices. In practice, we expect that the application-specific quantum algorithms are naturally resilient to quantum noise. In Ref. \cite{xiao2023practical}. we have found that QNN based GI is influenced by the quantum noise. For example, when increasing the quantum noise rate, the identification ability is decreased. In GI taks, the reconstruction ability is also influenced by the quantum noise rate. In current work, different from the quantum machine learning algorithm, the proposed algorithm does not require the datasets, which may be highly beneficial in practical scenarios. On drawback of QCSGI is its low-level generalization ability, which is a commonly demerit of CS-based optimization algorithms. The main point here is to showcase the robustness of QCSGI algorithm to quantum noise in the quantum circuit. 

When QNN is not perfect i.e. the fidelity of the quantum gates is smaller than 1, the performance of QCSGI under different quantum noise rates is presented in Fig. \ref{fig:noise_qc}. 
The detailed quantum noise simulation in quantum circuit can refer to Appendix \ref{Quconvolution}. From Fig. \ref{fig:noise_qc}, we can find that for different object, the reconstruction quality is nearly the same with the noise-free case especially at the late stage of the optimization. As QCSGI is different from deep learning GI, the quantum noise during optimization can be adaptive 'learned' to reconstruct the image. As long as the physical forward model is noise-free, the global optimum will always exist in the loss landscape. The quantum noise can slow down the optimization speed but as long as the iterations are large enough, the final recovered image will nearly have the same quality with noise-free case. Besides, we find that relatively larger noise rate does not decrease the reconstruction quality under 64, 128 and 256 measurements. In general, large noise rate in QNN will reduce the performance. However, in our hybrid model, QNN is further enhanced by the CNN model. The noisy quantum features can be further processed to obtain the recovered image. Under the strong supervision of physical-enhanced loss function, the noisy quantum features can be iteratively "filtered" or effectively error mitigated. However, in supervised QNN, the noisy quantum features is hard to be error-mitigated so that the performance will drop at large noise rate. In addition, in butterfly and wing object, the metrics of PSNR and SSIM with 512 measurements and quantum noise rate 0.005 are larger than other two cases, demonstrating that small noise rate is beneficial for QCSGI algorithm in large sampling ratio.

Additional results on other object are shown in Fig. \ref{fig:Add_noise}. Generally, sparse object such as the digit, plane has a better reconstruction under the same sampling ratio. QCSGI algorithm also shows the same results like other CS-based algorithms.

Overall, the reconstruction performance of QCSGI is not sensitive to quantum noise with the physical-enhanced loss function. Note that when the quantum noise is highly large, the expectation of observables may be highly random such that the convergence of our algorithm may be slow or even not converge. However, in current NISQ devices, the noise rate is controlled at a low level so that our algorithm is still appealing. Besides, we also find that large sampling ratio plays a much important role in the reconstruction performance.
\begin{table}[htbp]
	\centering
	\caption{Numerical results of IQP and Heisenberg encoding enabled QCSGI algorithm. Six objects are chosen to illustrate the reconstruction performance. PSNR denotes peak signal noise ratio and SSIM denotes structural similarity.}
	\label{tab2}
	\begin{tabular}{llllll}  
 	\hline
        \hline
		Measurements &  64   &  128 & 256 & 512 & 1024\\
		\hline
		IQP (SSIM, 2) & 0.83 & 0.92  & 0.94  & 0.97  & 0.95\\
		Hei. (SSIM, 2) & 0.86 & 0.92 & 0.94 & 0.95 & 0.95 \\
		IQP (PSNR, 2) & 21.79 & 25.25& 27.09& 30.11& 28.82\\
		Hei. (PSNR, 2) & 21.19& 24.81& 27.51& 28.62& 29.65 \\
		\hline
		IQP (SSIM, butterfly) & 0.27& 0.40& 0.56& 0.68& 0.79\\
		Hei. (SSIM, butterfly) & 0.30& 0.39& 0.55& 0.70& 0.78 \\
		IQP (PSNR, butterfly) & 14.56& 16.49& 18.41& 20.47& 22.04\\
		Hei. (PSNR, butterfly) & 14.57& 16.07& 18.41& 20.74& 22.06 \\
		\hline
		IQP (SSIM, epithelia) & 0.10& 0.13& 0.27& 0.52& 0.72\\
		Hei. (SSIM, epithelia) & 0.08& 0.13& 0.31& 0.52& 0.72 \\
		IQP (PSNR, epithelia) & 11.26&  11.36& 11.57& 13.53& 15.77\\
		Hei. (PSNR, epithelia) & 11.39& 11.47& 12.22& 13.62&  16.06 \\
		\hline
		IQP (SSIM, epithelia) & 0.10& 0.13& 0.27& 0.52& 0.72\\
		Hei. (SSIM, epithelia) & 0.08& 0.13& 0.31& 0.52& 0.72 \\
		IQP (PSNR, epithelia) & 11.26&  11.36& 11.57& 13.53& 15.77\\
		Hei. (PSNR, epithelia) & 11.39& 11.47& 12.22& 13.62&  16.06 \\
		\hline
		IQP (SSIM, mometasone) &0.31& 0.40& 0.47& 0.60&  0.69\\
		Hei. (SSIM, mometasone) &0.32& 0.37& 0.48& 0.60&  0.66 \\
		IQP (PSNR, mometasone) & 18.19& 20.37& 21.40& 23.16& 24.52\\
		Hei. (PSNR, mometasone) & 18.66& 19.74& 21.62& 23.44& 24.18 \\
			\hline
		IQP (SSIM, plane) &0.55& 0.61& 0.76&  0.80& 0.87\\
		Hei. (SSIM, plane) &0.49& 0.63& 0.75& 0.80& 0.85 \\
		IQP (PSNR, plane) & 18.23& 20.30& 23.92& 24.74& 24.67\\
		Hei. (PSNR, plane) & 17.30&  21.32& 23.46& 23.5& 24.22 \\
			\hline
		IQP (SSIM, wing) &0.15& 0.25& 0.44&  0.63& 0.75\\
		Hei. (SSIM, wing) &0.14& 0.26& 0.46& 0.59& 0.73 \\
		IQP (PSNR, wing) & 13.75& 14.18& 16.82& 18.95& 21.01\\
		Hei. (PSNR, wing) & 14.11& 13.97&  16.85& 18.60& 21.09\\
  	\hline
        \hline
	\end{tabular}
\end{table}

\subsection{Results of IQP and Heisenberg encoding}\label{I-H}

Here, we present the additional results with IQP and Heisenberg encoding strategies as discussed in Appendix \ref{app_qnn}. IQP and Heisenberg encoding strategy are more complex than angle encoding and also classically hard to simulate. From Table \ref{tab2}, we can find that IQP quantum circuit generally obtains slightly better performance in terms of the metrics of PSNR and SSIM. 
\begin{figure*}[ht]
    \centering
    \includegraphics[width=0.8\textwidth]{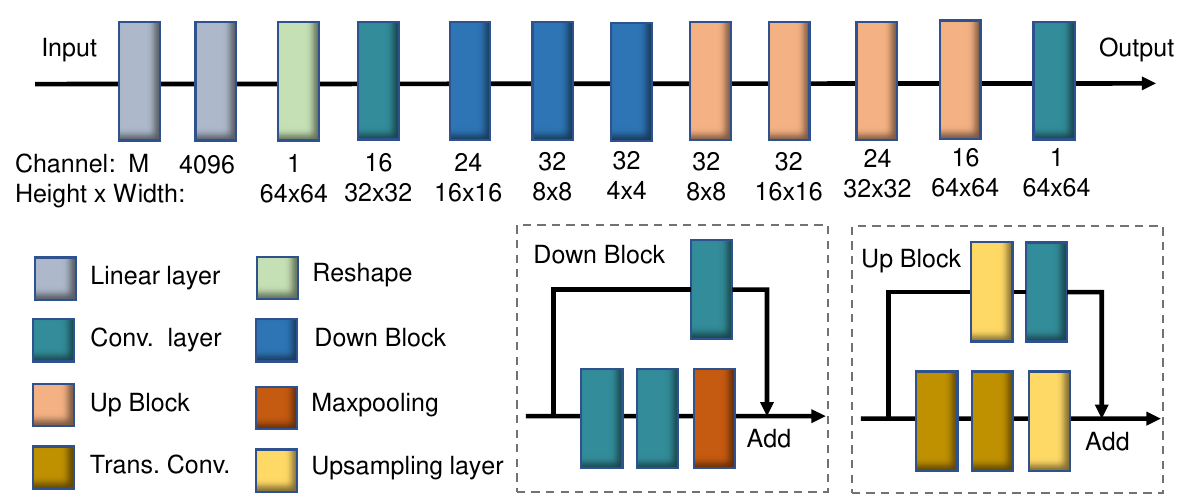}
    \caption{Classical convolutional neural network (CNN) architecture used in this work. }
    \label{fig:network}
\end{figure*}

\section{Hyperparameters specification}
We conduct Gaussian initialization of QNN training. Gaussian initialization generates the weight values from Gaussian probability distributions. Specifically, similar to the typical Xavier initialization, we make use of the geometrical structure i.e. the width of QNN $N_Q$ to initialize the network parameters. Formally, 
\begin{equation}
    \bm{\theta} \sim \mathcal{N}(0, N_Q), \, f(\theta_i) = \frac{1}{\sqrt{2\pi N_Q}}\exp\left(-\frac{\theta_i^2}{2N_Q}\right)
\end{equation}
where $\theta_i$ is the $i$th weight of QNN. Moreover, according to the trick of \cite{zhang2022escaping}, the initial weight parameter of QNN should be concentrated in a small area such that we multiple the random weight by a small scalar 0.1 or 0.01 during the optimization process. The weighted factor $\mu$ is set as $10^{-6}$.

\begin{table}[h]\color{black}
\centering
\caption{\textcolor{black}{The number of parameters used in QNNs and CNNs. The bold digits denotes the the distinguished part of the QNN and CNN. As shown in Fig.~\ref{fig:network}, the first input layer of the QNN is different from the CNN. The subsequent layers and structures are the same.}}

\label{tab:parameters}
\begin{tabular}{ccccc}
\hline \hline 
\multicolumn{1}{l}{} & \multicolumn{2}{c}{QCSGI}                                                                                      & \multicolumn{2}{c}{CNN}                                                                           \\ \hline
\multicolumn{1}{l}{} & \begin{tabular}[c]{@{}c@{}}Quan. \\ para.\end{tabular} & \begin{tabular}[c]{@{}c@{}}Class. \\ para.\end{tabular} & \begin{tabular}[c]{@{}c@{}}Substit. \\ param. \\ (Linear layer)\end{tabular} & \begin{tabular}[c]{@{}c@{}}Class. \\ para.\end{tabular} \\ \hline
64                   & \textbf{1,020}                                                & 333,281                                                    & \textbf{4,160}                                                                         & 333,281                                                    \\
128                  & \textbf{2,044}                                                & 595,425                                                    & \textbf{16,512}                                                                        & 595,425                                                    \\
256                  & \textbf{4,092}                                                & 1,119,713                                                   & \textbf{65,792}                                                                        & 1,119,713                                                   \\
512                  & \textbf{8,188}                                                & 2,168,289                                                   & \textbf{262,656}                                                                       & 2,168,289                                                   \\
1024                 & \textbf{16,380}                                               & 4,265,441                                                   & \textbf{1,049,600}                                                                     & 4,265,441                                                   \\ \hline\hline 
\end{tabular}
\end{table}

As demonstrated in the Method section, the hybrid model consists of the quantum part which is given in the former section and the classical part. And we compare the proposed model with classical convolution neural network (CNN) model. For fair comparison, we actually replace the quantum part of the proposed model with a classical linear layer in order to serve as the classical CNN model used for comparing. That is, when comparing a hybrid quantum network with a classical network, the quantum part of the network structure in Fig. \ref{fig:1} is replaced by a classical linear layer. At this point the two methods have similar number of parameters (not strictly equal). The classical CNN architecture is depicted in Fig. \ref{fig:network} and the details of number of parameters are listed in Table \ref{tab:parameters}. The 1D bucket input with the length of $M$ first operates by a linear layer with Leaky ReLU activation, which acts as the replacement of quantum part.  And then, the other linear layer and the reshape operation increase its dimension to 64$\times$64 with one channel. After a convolution layer and the down block, the feature is down-sampled to 4$\times$4 with 32 channels. Finally, the up block and a a convolution layer operate it to 64$\times$64 as the output. The output size of each layer or block is marked in Fig. \ref{fig:network}. The up block and the down block are based on the residual network, and the detailed structure of two blocks are also depicted. The activation function after each layer is leaky ReLU \cite{arora2016understanding} with the factor of 0.2. The optimizer is Adam with the learning rate of 0.05. The algorithm runs on a computer with an Intel CPU I7-6800K, 64GB RAM, and two NVIDIA Titan Xp GPU. The deep-learning platform is Pytorch and Tensorflow.
The numerical simulation of quantum neural network is based on the open-source framework called TensorCircuit \cite{zhang2023tensorcircuit}, which supports Just-in-time compiling and Vmap operation to accelerates the training of quantum neural networks. Quantum neural network is deployed in the QSIP server which contains 192 CPUs and 2 RTX 3090 GPUs each with 24 GB memory.
\end{appendix}

\end{document}